\documentclass[10pt,onecolumn]{iontran}
\usepackage[pdftex,bookmarksopen=true,bookmarks=true]{hyperref}
\usepackage{bookmark}
\usepackage{subfigure,wrapfig,graphicx,booktabs,fancyhdr,amsmath,amsfonts}
\usepackage{cite,bm,amssymb,amsthm,url,multirow,times,enumitem,comment}

\usepackage[linesnumbered,ruled]{algorithm2e}
\usepackage[table]{xcolor}
\usepackage[skip=0.333\baselineskip]{caption}
\usepackage{array}
\usepackage{float,xspace,siunitx}
\usepackage{pdfpages,pdflscape}

\newcommand{\vb}{\boldsymbol}

\newcommand{\vbc}[1]{\check{\boldsymbol{#1}}}
\newcommand{\vbb}[1]{\bar{\boldsymbol{#1}}}
\newcommand{\vbt}[1]{\tilde{\boldsymbol{#1}}}

\newcommand{\tr}{\mathsf{T}}

\newcommand{\ba}{\begin{array}}
\newcommand{\ea}{\end{array}}

\renewcommand{\equiv}{\triangleq}

\newcommand{\mr}[1]{{\mathrm{#1}}}
\newcommand{\norm}[1]{\left\lVert#1\right\rVert}

\newcommand{\chol}[1]{\texttt{chol}\left({#1}\right)}

\begin{document}

\setlength\parindent{0pt}
\setlength\parskip{6pt}
\title{\huge \textbf{Carrier-phase and IMU based GNSS Spoofing Detection for Ground Vehicles}}
\author{\normalsize Zachary Clements, James E. Yoder, and Todd E. Humphreys \\
  \emph{Radionavigation Laboratory} \\
  \emph{The University of Texas at Austin} \\ \vspace{5mm}
}

\maketitle

\newif\ifpreprint
\preprinttrue

\ifpreprint

\pagestyle{plain}
\thispagestyle{fancy}  
\fancyhf{} 
\renewcommand{\headrulewidth}{0pt}
\title{\huge \textbf{Carrier-phase and IMU based GNSS Spoofing Detection for Ground Vehicles}}
\rfoot{\footnotesize \bf Preprint of the 2022 ION ITM Conference\\
	Long Beach, CA, Jan. 25-27, 2022} \lfoot{\footnotesize \bf
	Copyright \copyright~2022 by Zachary Clements, \\ James E. Yoder, and Todd
	E. Humphreys}
\else

\thispagestyle{empty}
\pagestyle{empty}

\fi

\section*{Biographies}
Zachary Clements (BS, Electrical Engineering, Clemson University) is a graduate
student in the department of Aerospace Engineering and Engineering Mechanics at
The University of Texas at Austin, and a member of the UT Radionavigation
Laboratory. His research interests include GNSS signal processing, spoofing
detection techniques, software-defined radio, and sensor fusion.

James Yoder (BS, Electrical Engineering, The University of Texas at Austin) is
a graduate student in the department of Aerospace Engineering and Engineering
Mechanics at The University of Texas at Austin, and a member of the UT
Radionavigation Laboratory.  His research interests include GNSS signal
processing, estimation, and sensor fusion techniques for centimeter-accurate
navigation.

Todd Humphreys (BS, MS, Electrical Engineering, Utah State University; PhD,
Aerospace Engineering, Cornell University) is a professor in the department of
Aerospace Engineering and Engineering Mechanics at The University of Texas at
Austin, where he directs the Radionavigation Laboratory.  He specializes in the
application of optimal detection and estimation techniques to problems in
secure, collaborative, and high-integrity perception, with an emphasis on
navigation, collision avoidance, and precise timing.  His awards include The
University of Texas Regents' Outstanding Teaching Award (2012), the National
Science Foundation CAREER Award (2015), the Institute of Navigation Thurlow
Award (2015), the Qualcomm Innovation Fellowship (2017), the Walter Fried Award
(2012, 2018), and the Presidential Early Career Award for Scientists and
Engineers (PECASE, 2019). He is a Fellow of the Institute of Navigation and of
the Royal Institute of Navigation.

\section*{Abstract}
This paper develops, implements, and validates a powerful single-antenna
carrier-phase-based test to detect Global Navigation Satellite Systems (GNSS)
spoofing attacks on ground vehicles equipped with a low-cost
inertial measurement unit (IMU).  Increasingly-automated ground vehicles
require precise positioning that is resilient to unusual natural or accidental
events and secure against deliberate attack.  This paper's spoofing detection
technique capitalizes on the carrier-phase fixed-ambiguity residual cost produced 
by a well-calibrated carrier-phase-differential GNSS (CDGNSS) estimator that is
tightly coupled with a low-cost IMU.  The carrier-phase fixed-ambiguity residual cost 
is sensitive at the sub-centimeter-level to discrepancies between measured carrier phase 
values and the values predicted by prior measurements and by the dynamics model, which is
based on IMU measurements and on vehicle constraints.  Such discrepancies will
arise in a spoofing attack due to the attacker's practical inability
to predict the centimeter-amplitude vehicle movement caused by roadway
irregularities.  The effectiveness of the developed spoofing detection method
is evaluated with data captured by a vehicle-mounted sensor suite in Austin,
Texas.  The dataset includes both consumer- and industrial-grade IMU data and a
diverse set of multipath environments (open sky, shallow urban, and deep urban).
Artificial worst-case spoofing attacks injected into the dataset are
detected within two seconds.

\section*{Introduction}
The combination of easily-accessible low-cost Global Navigation Satellite
System (GNSS) spoofers and the emergence of increasingly-automated GNSS-reliant
ground vehicles prompts a need for fast and reliable GNSS spoofing detection
\cite{volpe2001gps, psiakiNewBlueBookspoofing}. To underscore this point,
Regulus Cyber recently spoofed a Telsa Model 3 on autopilot mode, causing the
vehicle to suddenly slow and unexpectedly veer off the main road
\cite{mit2020TeslaSpoof}.

Among GNSS signal authentication techniques, signal-quality-monitoring (SQM)
and multi-antenna could be considered for implementation on ground vehicles
\cite{psiaki2016gnssSpoofing}.  However, SQM tends to perform poorly on dynamic
platforms in urban areas where strong multipath and in-band noise are common
\cite{wesson2018pincer,gross2018maximum,psiaki2016gnssSpoofing,humphreysGNSShandbook},
and multi-antenna spoofing detection techniques, while effective
\cite{montgomery2009pcgs,psiaki2014wrod}, are disfavored by automotive
manufacturers seeking to reduce vehicle cost and aerodynamic drag.  Thus, there
is a need for a single-antenna GNSS spoofing detection technique that performs
well on ground vehicles despite the adverse signal-propagation conditions in an
urban environment.

In a concurrent trend, increasingly-automated ground vehicles demand
ever-stricter lateral positioning to ensure safety of operation.  An
influential recent study calls for lateral positioning better than 20 cm on
freeways and better than 10 cm on local streets (both at 95\%)
\cite{reid2019localization}.  Such stringent requirements can be met by
referencing lidar and camera measurements to a local high-definition map
\cite{ye2019tightly,chiang2020navigation}, but poor weather (heavy rain, dense
fog, or snowy whiteout) can render this technique unavailable
\cite{narula2021radarpositioningjournal}.  On the other hand, recent progress
in precise (dm-level) GNSS-based ground vehicle positioning, which is
impervious to poor weather, has demonstrated surprisingly high (above 97\%)
solution availability in urban areas \cite{yoder2022tightCoupling}.  This
technique is based on carrier-phase differential GNSS (CDGNSS) positioning,
which exploits GNSS carrier phase measurements having mm-level precision but
integer-wavelength ambiguities \cite{teunissenGNSShandbookCdgnss}.

Key to the promising results in \cite{yoder2022tightCoupling} is the tight
coupling of CDGNSS and IMU measurements, without which high-accuracy CDGNSS
solution availability is significantly reduced due to pervasive signal blockage
and multipath in urban areas (compare the improved performance of
\cite{yoder2022tightCoupling} relative to \cite{humphreys2019deepUrbanIts}).
Tight coupling brings mm-precise GNSS carrier phase measurements into
correspondence with high-sensitivity and high-frequency inertial sensing.  The
particular estimation architecture of \cite{yoder2022tightCoupling}
incorporates inertial sensing via model replacement, in which the estimator's
propagation step relies on bias-compensated acceleration and angular rate
measurements from the IMU instead of a vehicle dynamics model.  As a
consequence, at each measurement update, an \emph{a priori} antenna position is
available whose delta from the previous measurement update accounts for all
vehicle motion sensed by the IMU, including small-amplitude high-frequency
motion caused by road irregularities.  Remarkably, when tracking authentic GNSS
signals in a clean (open sky) environment, the GNSS carrier phase predicted by
the \emph{a priori} antenna position and the actual measured carrier phase
agree to within millimeters.

This paper pursues a novel GNSS spoofing detection technique based on a simple
but consequential observation: it is practically impossible for a spoofer to
create a false ensemble of GNSS signals whose carrier phase variations, when
received through the antenna of a target ground vehicle, track the phase values
predicted by inertial sensing.  In other words, antenna motion caused by road
irregularities, or rapid braking, steering, etc., is sensed with high fidelity
by an onboard IMU but is unpredictable at the sub-cm-level by a would-be
spoofer.  Therefore, the differences between IMU-predicted and measured carrier
phase values offer the basis for an exquisitely sensitive GNSS spoofing detection 
statistic.  What is more, such carrier phase fixed-ambiguity residual cost is generated 
as a by-product of tightly-coupled inertial-CDGNSS vehicle position estimation such 
as performed in \cite{yoder2022tightCoupling}.

Two difficulties complicate the use of fixed-ambiguity residual cost 
for spoofing detection.  First is the integer-ambiguous nature of the carrier phase
measurement \cite{teunissenGNSShandbookCdgnss}, which causes the
post-integer-fix residual cost to equal not the difference
between the measured and predicted carrier phase, as would be the case for a
typical residual, but rather this difference modulo an integer number of
carrier wavelengths.  Such integer folding complicates development of a
probability distribution for a detection test statistic based on carrier phase
fixed-ambiguity residual cost.

Second, the severe signal multipath conditions in urban areas create thick
tails in any detection statistic based on carrier phase measurements.  Setting a
detection threshold high enough to avoid false spoofing alarms caused by mere
multipath could render the detection test insensitive to dangerous forms of
spoofing.  Reducing false alarms by accurately modeling the effect of a
particular urban multipath environment on the detection statistic would be a
Sisyphean undertaking, requiring exceptionally accurate up-to-date 3D models of
the urban landscape, including materials properties.

This paper takes an empirical approach to these difficulties.  It does not
attempt to develop a theoretical model to delineate the effects of integer
folding or multipath on its proposed carrier-phase fixed-ambiguity residual cost based detection
statistic. Rather, it develops null-hypothesis empirical distributions for the
statistic in both shallow and deep urban areas, and uses these distributions to
demonstrate that high-sensitivity spoofing detection is possible despite
integer folding and urban multipath.

\subsection*{Related Work}
The idea of using coupled GNSS and inertial sensing to detect GNSS spoofing was
first explored for aviation
\cite{khanafseh2014raim,tanil2017detecting,tanil2018insMonitor,
  tanil2018experimental,tanil2018sequential,kujur2020solution}.  Wind gusts and
turbulence cause rapid movement of aircraft that are instantaneously reflected
in calibrated inertial measurements.  As with road irregularities for ground
vehicles, a GNSS spoofer will find it challenging to track and replicate such
movements in real-time.  However, this prior work either did not exploit
carrier-phase measurements or relied on a tactical-grade IMU, rendering
solutions either too slow (long time-to-detect) or too expensive.

Accumulating innovation faults within a specified time window in a loosely
coupled INS/GNSS Kalman filter was investigated in \cite{liu2019analysis}.  For
a given time window, there are two ways to accumulate the slowly-drifting
faults. One averages the normalized sum-squared innovations of each epoch
(innovation averaging); the other averages the measurements within a time
window and subsequently performs a snapshot test (measurement averaging).
Innovation averaging has little effect on the Kalman filter prediction and
filtering process, so it can be easier to deploy and can be designed as an
add-on function.  Measurement averaging requires small modifications of the
Kalman filter measurement update process.  The position-domain spoofing
detection strategy in \cite{liu2019analysis} required 15 seconds to detect a
fairly obvious spoofing attack with position drift of 5 m/s.  Such a time to
detect is unacceptably long for an automated ground vehicle.

Prior work in spoofing detection specifically for ground vehicles demonstrated
that low-cost IMUs could be used to detect GNSS spoofing by constructing a
coherency test between the GNSS and inertial measurements \cite{curran2017use}.
But the test statistic in \cite{curran2017use} was constructed from
position-domain measurements, and so is much less sensitive than the
carrier-phase-based test proposed in the current paper, resulting in
an unacceptably-long (3 minute) time to detection.

\subsection*{Contributions}
This paper's primary contributions are (i) the development and verification of
a highly sensitive all-environment single-antenna GNSS spoofing detection
technique based on carrier-phase fixed-ambiguity residual cost produced by a
well-calibrated CDGNSS solution that is tightly coupled with a low-cost IMU,
(ii) the introduction of an artificial worst-case spoofing methodology, and
(iii) a comparison between industrial- and consumer-grade IMUs for spoofing
detection within the proposed framework.

\section*{Measurement Model}  
The full formulation of the measurement model for the tightly-coupled GNSS-IMU
estimator on which this paper's spoofing detection technique is based may be
found in \cite{yoder2022tightCoupling}. Key developments are presented here for
the reader's convenience.

The estimator ingests $N_k$ pairs of double-difference (DD) GNSS
observables at each GNSS measurement epoch, with each pair composed of a
pseudorange and a carrier phase measurement. The measurement vector at epoch
$k$ is
\begin{align*}
	\vb z_{k} \triangleq \left[	\vb \rho_{k}^\tr, \vb \phi_{k}^\tr	\right]^\tr \in \mathbb{R}^{2N_k}
\end{align*}
where $\vb \rho_{k}$ and $\vb \phi_{k}$ are vectors of double-difference
pseudorange and carrier phase measurements, both in meters.  At epoch $k$,
after linearizing about the \emph{a priori} state estimate, a measurement model
can be expressed as 
\begin{equation}
	\label{eq:innovations}
	\vb \nu_{\mr k} = \vb H_{\mr rk}\delta \vb x_k - \vb H_{\mr nk} \vb n_k + \vb w_{\nu k}, \quad \vb w_{\nu} \sim \mathcal{N}(\vb 0,\,\Sigma_k)
\end{equation}
where $\vb \nu_{\mr k}$ is the difference between the measurement $\vb z_{k}$
and its modeled value based on the \emph{a priori} state estimate,
$\vb H_{\mr rk}$ and $\vb H_{\mr nk}$ are Jacobians, $\delta \vb x_k$ is the
state estimate error vector, $\vb n_k \in \mathbb{Z}^{N_k}$ is the the integer
ambiguity vector, and $\vb w_{\nu k}$ is noise.  A short-baseline regime is
assumed for the DD measurements, which implies that ionospheric, tropospheric,
ephemeris, and clock errors are cancelled in the double differencing, leaving
$\vb w_{\nu k}$ to account only for multipath and receiver thermal noise.  The
prior on the real-valued error state can be expressed in terms of the following
data equation:
\begin{equation}
	\label{eq:x_cost}
	\vb 0 = \delta \vb x_k + \vb w_{x k}, \quad \vb w_{x} \sim \mathcal{N}(\vb 0,\, \vbb P_k)
\end{equation}

The CDGNSS measurement update of the tightly-coupled GNSS-IMU estimator can be
cast in square-root form for greater numerical robustness and algorithmic
clarity \cite{psiaki2005relative}.  Given $\vb \nu_{ k}$, $\vb H_{\rm rk}$,
$\vb H_{\rm nk}$, and the data equation above, the measurement update can be
defined as the process of finding $\delta \vb x_k$ and $\vb n_k$ to minimize
the cost function
\begin{align*}
	J_k(\delta \vb x_k, \vb n_k) =
	\norm{
		\vb \nu_{ k} 
		- \vb H_{\mr rk}\delta \vb x_k 
		- \vb H_{\mr nk} \vb n_k
	}^2_{\vb \Sigma_{k}^{-1}} +
	\norm{
		\vb \delta x_k
	}^2_{\vbb P_k^{-1}}
\end{align*}
The vector cost components can be normalized by left multiplying with
square-root information matrices based on Cholesky factorization
$\vb R_{\mr k} = \chol{\vb \Sigma_{k}{}^{-1}}$,
$\vbb R_{xxk} = \chol{\vbb P_k^{-1}}$:
\begin{align*}
	J_k(\delta \vb x_k, \vb n_k)  = &\norm{
		\begin{bmatrix}
			\vb 0 \\ \vb R_{\mr k} \vb \nu_{ k}
		\end{bmatrix} -
		\begin{bmatrix}
			\vbb R_{xxk} \\ \vb R_{\mr k} \vb H_{\mr rk} 
		\end{bmatrix}
		\delta \vb x_k -
		\begin{bmatrix}
			\vb 0 \\ \vb R_{k} \vb H_{\mr rk}
		\end{bmatrix}
		\vb n_k
	}^2 \\ 
	= &\norm{
		\vb \nu_k' - 
		\begin{bmatrix}
			\vb H_{\mr rk}' \\
			\vb H_{\mr nk}'
		\end{bmatrix}
		\begin{bmatrix}
			\delta \vb x_k \\ \vb n_k
		\end{bmatrix}
	}^2
\end{align*}
The cost $J_k$ can be decomposed via QR factorization
\begin{align*}
	\left[\vbt {Q}_k, \vbt {R}_k \right] = \mathtt{qr}\left(
	\begin{bmatrix}
		\vb H_{\mr rk}' \\
		\vb H_{\mr nk}'
	\end{bmatrix}
	\right)
\end{align*}
where matrix $\vbt {Q}_k$ is orthogonal and $\vbt {R}_k$ is upper
triangular.  Because $\vbt {Q}_k$ is orthogonal, the components of $J_k$
inside the norm can be left-multiplied by $\vbt {Q}_k^\tr$ without
changing the cost, and $J_k$ can be decomposed into 3 terms:
\begin{align}
	\label{eq:sr_cost} 
	J_k(\delta \vb x_k, \vb n_k)  =  & \norm{
		\vbt Q_k^\tr \vb \nu'_k -
		\vbt R_k 
		\begin{bmatrix}
			\delta \vb x_k  \\ \vb n_k
		\end{bmatrix}	
	}^2 \nonumber \\ 
	= & \norm {
		\begin{bmatrix}
			\vb \nu''_{1k} \\ \vb \nu''_{2k} \\ \vb \nu''_{3k}
		\end{bmatrix} - 
		\begin{bmatrix}
			\vb R_{xxk} & \vb R_{xnk} \\
			\vb 0      & \vb R_{nnk} \\
			\vb 0      & \vb 0
		\end{bmatrix}
		\begin{bmatrix}
			\delta \vb x_k \\ \vb n_k
		\end{bmatrix}
	}^2  \\  
	= & \underbrace{\norm{
			\vb \nu_{1k}'' 
			- \vb R_{xxk} \delta \vb x_k 
			- \vb R_{xnk} \vb n_k
		}^2}_{J_{1k}\left(\delta\vb x_k, \vb n_k\right)} 
	+ \underbrace{\norm{ 
			\vb \nu_{2k}'' - \vb R_{nnk} \vb n_k				
		}^2}_{J_{2k}(\vb n_k)} \nonumber 
	 + \underbrace{\norm {\vb \nu_{3k}''}^2}_{J_{3k}} \nonumber
\end{align}
If both the measurement model and $\vbb R_{xxk}$ are not ill-conditioned, then
$\vb R_{xxk}$ and $\vb R_{nnk}$ are invertible.  $J_{1k}$ can be zeroed for any
value of $\vb n_k$ due to the invertibility of $\vb R_{xxk}$.  $J_{3k}$ is the
irreducible cost, and, under a single-epoch ambiguity resolution scheme, can be
shown to be equal to the normalized innovations squared (NIS) associated with
the double-difference pseudorange measurements.

$J_{2k}$ is the extra cost incurred by enforcing the integer constraint on
$\vb n_k$.  If $\vb n_k$ is allowed to take any real value (the \emph{float
  solution}), $J_{2k}$ can be zeroed due to the invertibility of $\vb R_{nnk}$.
The float solution $\{\delta \vbt x_k, \vbt n_k\}$ is formed by choosing
$\delta \vbt x_k$ and $\vbt n_k$ to zero $J_{1k}$ and $J_{2k}$. Because
$\vbt R_k$ is upper triangular, these values can be found by efficient
backsubstitution. The \emph{fixed solution} $\{\delta \vbc x_k,\vbc n_k\}$ is
found via an integer least squares (ILS) solver, yielding
\begin{equation}
	\label{eq:ils_solve}
	\begin{aligned}
		\vbc n_k = {}& \arg \underset{\vb n_k \in \mathbb{Z}^{N_k}}{\min} \, J_{2k}(\vb n_k) \\
		\delta \vbc x_k = {}& \vb R_{xxk}^{-1} \left( \vb \nu_{1k}'' - \vb R_{xnk} \vbc n_k \right)
	\end{aligned}
\end{equation}	
Note that $\vb R_{xxk}$ is the \emph{a posteriori} state vector square-root
information matrix conditioned on $\vb n_k = \vbc n_k$.

\section*{Test Statistic}
Key to this paper's spoofing detection statistic is the integer-fixed
carrier-phase residual cost
\[\epsilon_{\phi k} = J_{2k}(\vbc n_k)\]  which can also be thought of as
the ILS solution cost \cite{psiakiAmbiguity2007}.  This is small whenever the
carrier phase measurements are consistent with the prior state estimate, the
pseudorange measurements, and with the assumption of integer-valued
carrier-phase ambiguities.  It is one of several acceptance test statistics
used to decide whether the fixed solution $\vbc n_k$ is correct with high
probability \cite{teunissenGNSShandbookCdgnss}. In \cite{s_mohiuddin07_wia},
$\epsilon_{\phi k}$ was incorporated in a statistic used to detect carrier
cycle slips.  It can similarly be used to detect false integer fixes, just as
with other integer aperture acceptance test statistics, or the lingering
effects of conditioning the real-valued part of the state $\delta \vb x_k$ on a
previous false fix \cite{yoder2022tightCoupling}.

Furthermore, $\epsilon_{\phi k}$ provides is a highly sensitive statistic for
spoofing detection.  When no spoofing is present, there is tight agreement
between the IMU-propagated \emph{a priori} state estimate and GNSS data
resulting in a small $\epsilon_{\phi k}$. If the vehicle hits a bump in the
road, the GNSS antenna phase center will rise by a few centimeters, and the
inertial sensor will detect a corresponding acceleration, which will get
propagated through to the \emph{a priori} state. On the other hand, when
spoofing is present, a discrepancy between inertial and GNSS data will arise at
the carrier-phase level, leading to $\epsilon_{\phi k}$ being larger than
usual.

A windowed sum of $\epsilon_{\phi k}$ offers even greater sensitivity to false
fix events at the expense of a longer time to detect.  The test statistic used
to detect spoofing in this paper is the \emph{windowed fixed-ambiguity residual 
cost} (WFARC), $\Psi_k$.  This is calculated over a moving window of fixed
length $l$ of past GNSS measurement epochs. It has $N_{\Psi_k}$ degrees of
freedom and is calculated by
\begin{align*}
	\Psi_k      \equiv \sum_{n=k-l+1}^{k} \epsilon_{\phi n}, \quad
	N_{\Psi_k}  \equiv \sum_{n=k-l+1}^{k} N_{n}
\end{align*}
where $N_k$ is the number of DD carrier phase measurements at epoch $k$.  In
this paper, a window length of $l$ = 10 past GNSS measurement epochs (amounting
to a window of 2 seconds) is used.

If the filter is consistent and the integer ambiguities are correctly resolved,
then $\Psi_k$ should be approximately $\chi^2$-distributed with $N_{\Psi_k}$
degrees of freedom. This distribution is approximate due to the
``integer-folding'' effect: large phase residuals are not possible because of
integer-cycle phase wrapping.  A statistical consistency test can be performed
by choosing a desired false-alarm rate $\bar P_{f,\Psi}$ and declaring a false
fix if $\Psi_k > \gamma_{\Psi k}$, where the threshold $\gamma_{\Psi k}$ is
calculated by evaluating the inverse cumulative distribution function of
$\chi^2(N_{\Psi_k})$ at $\bar P_{f,\Psi}$.

Compared to a single residual $\epsilon_{\phi k}$, $\Psi_k$ has greater
statistical power for consistency testing and helps avoid premature declaration
of spoofing due to sporadic measurement outliers. However, increasing the
window length $l$ also increases the latency to detect a spoofing event.  This
statistical test is conducted at each GNSS measurement epoch.  The null
hypothesis, $H_0$ denotes no spoofing detected, and the alternate hypothesis,
$H_1$ indicates the detection of spoofing.  These are declared according to the rule
\begin{align*}
	\delta(\Psi_k) = \left\{ \ba{ccc} H_0 & \mbox{if} & \Psi_k <
                                                           \gamma_{\Psi k}\\
                        H_1 & \mbox{if} & \Psi_k \geq \gamma_{\Psi k} \ea
                                          \right.
\end{align*}

\section*{Data Collection}
\begin{figure}[H]
	\centering
	\begin{minipage}{.5\textwidth}
		\centering
		\scalebox{-1}[1]{\includegraphics[width=1\linewidth]{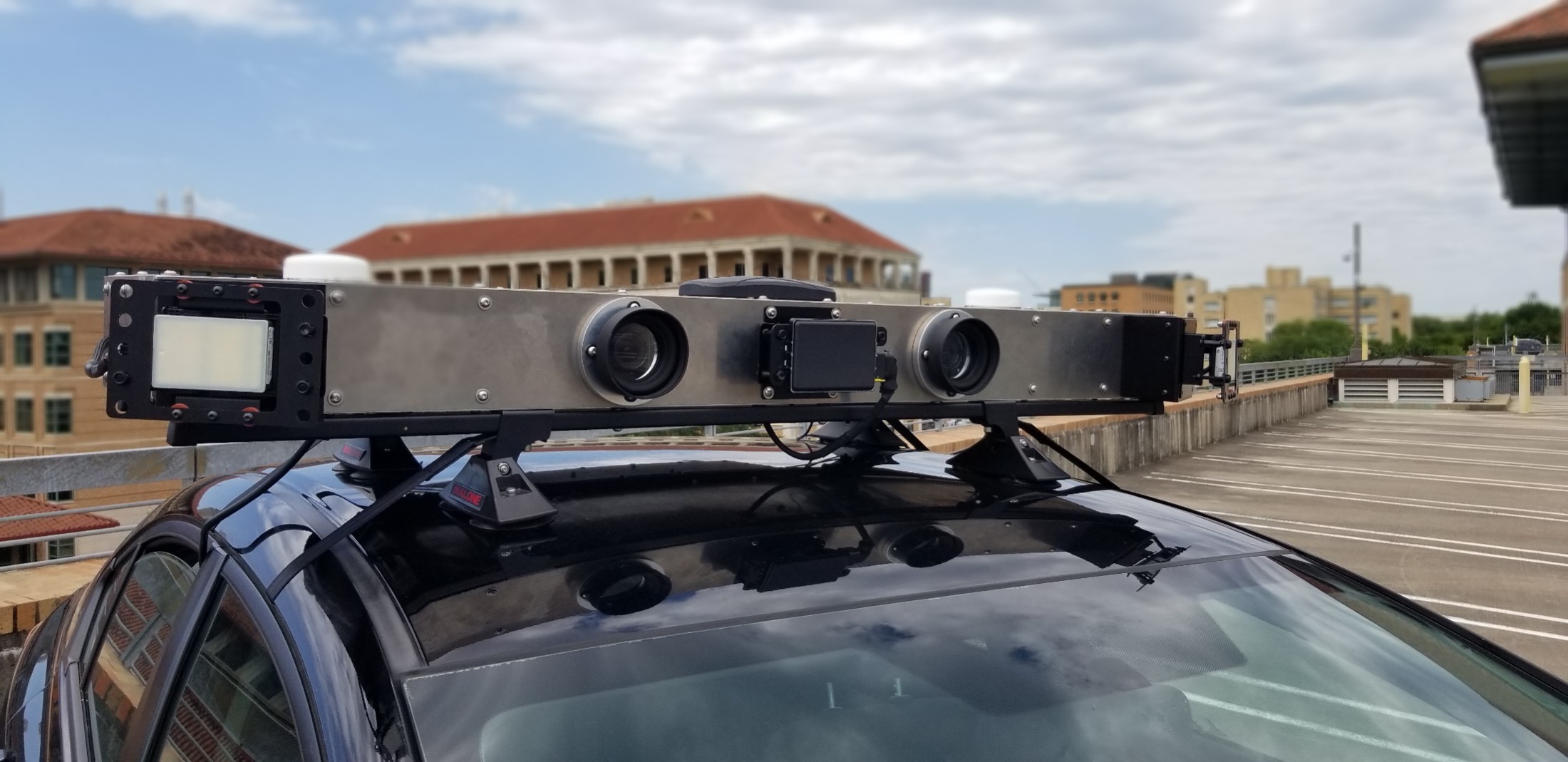}}
	\end{minipage}%
	\begin{minipage}{.5\textwidth}
		\centering
		\includegraphics[width=1\linewidth]{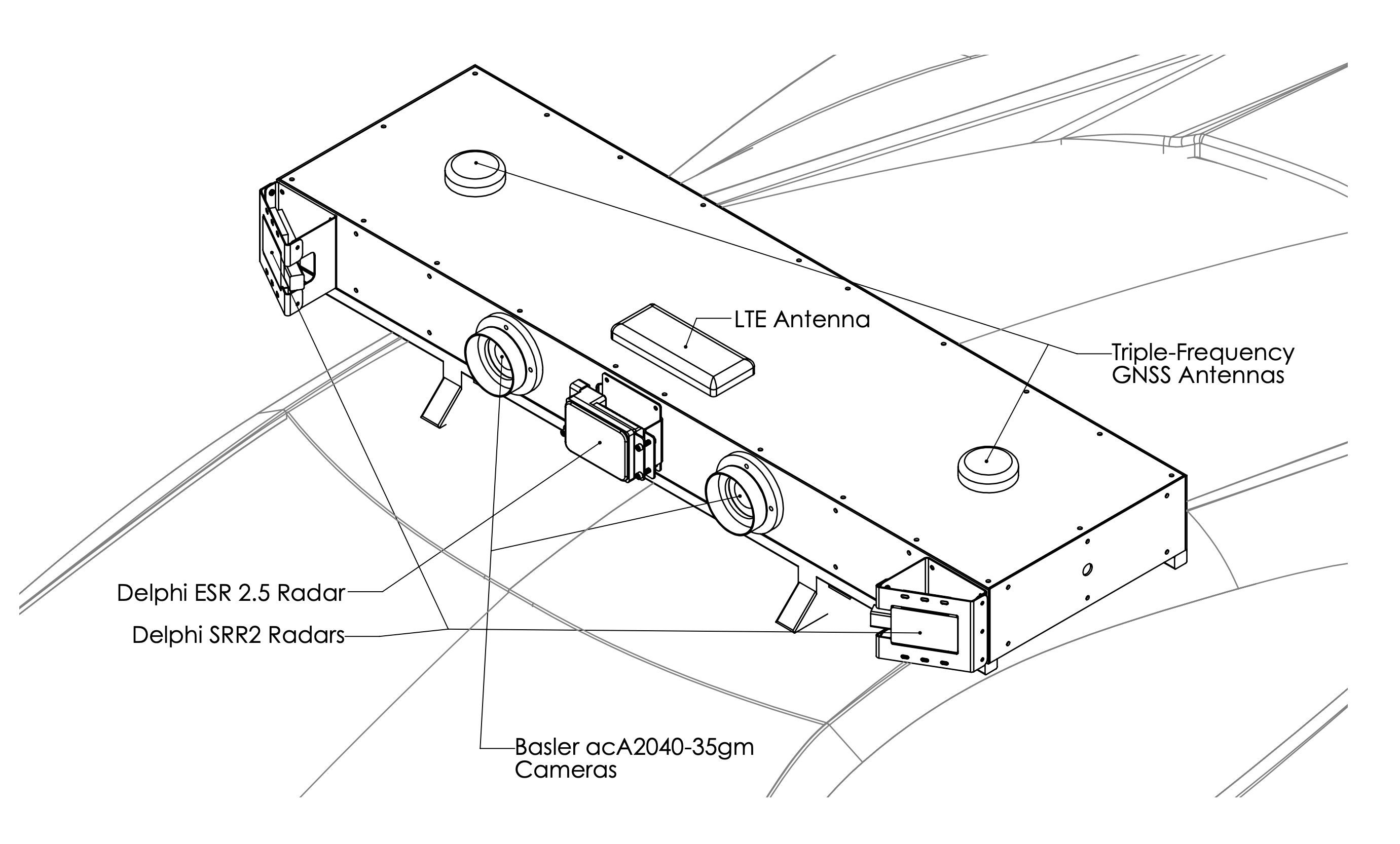}
	\end{minipage}
	\caption{The UT RNL has developed a multi-modal ground-vehicle-mounted integrated perception platform call the \emph{Sensorium}. It houses three different types of IMU, two triple-frequency GNSS antennas, three radar sensors, and two cameras. An extensive localization pipeline based on these sensors has been developed.}
	\label{fig:sensorium}
\end{figure}

\begin{figure*}[h!]
	\centering
	\begin{minipage}[b]{0.325\textwidth}
		\centering
		\includegraphics[width=\linewidth]{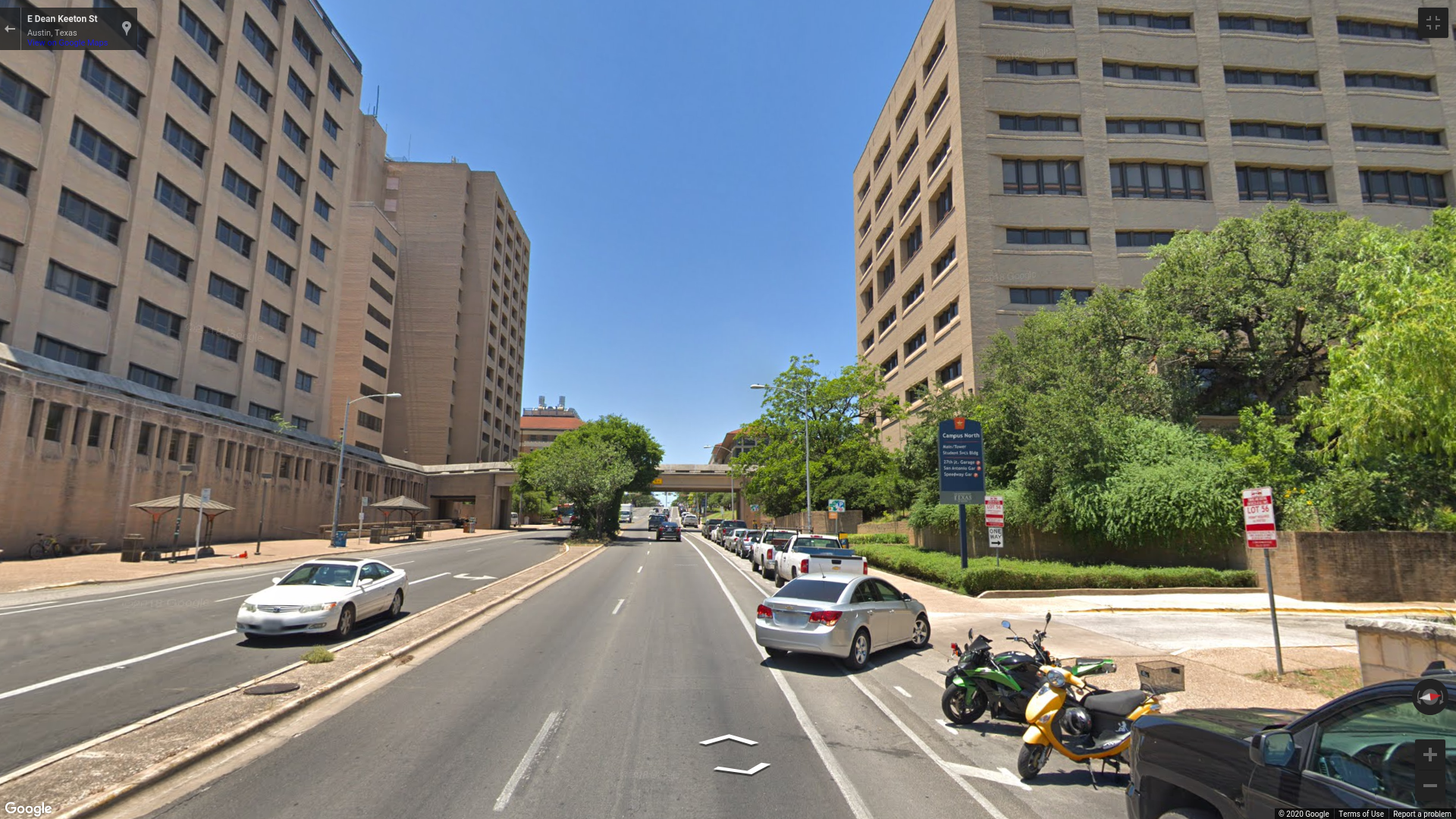}
	\end{minipage}
	\begin{minipage}[b]{0.325\textwidth}
		\centering
		\includegraphics[width=\linewidth]{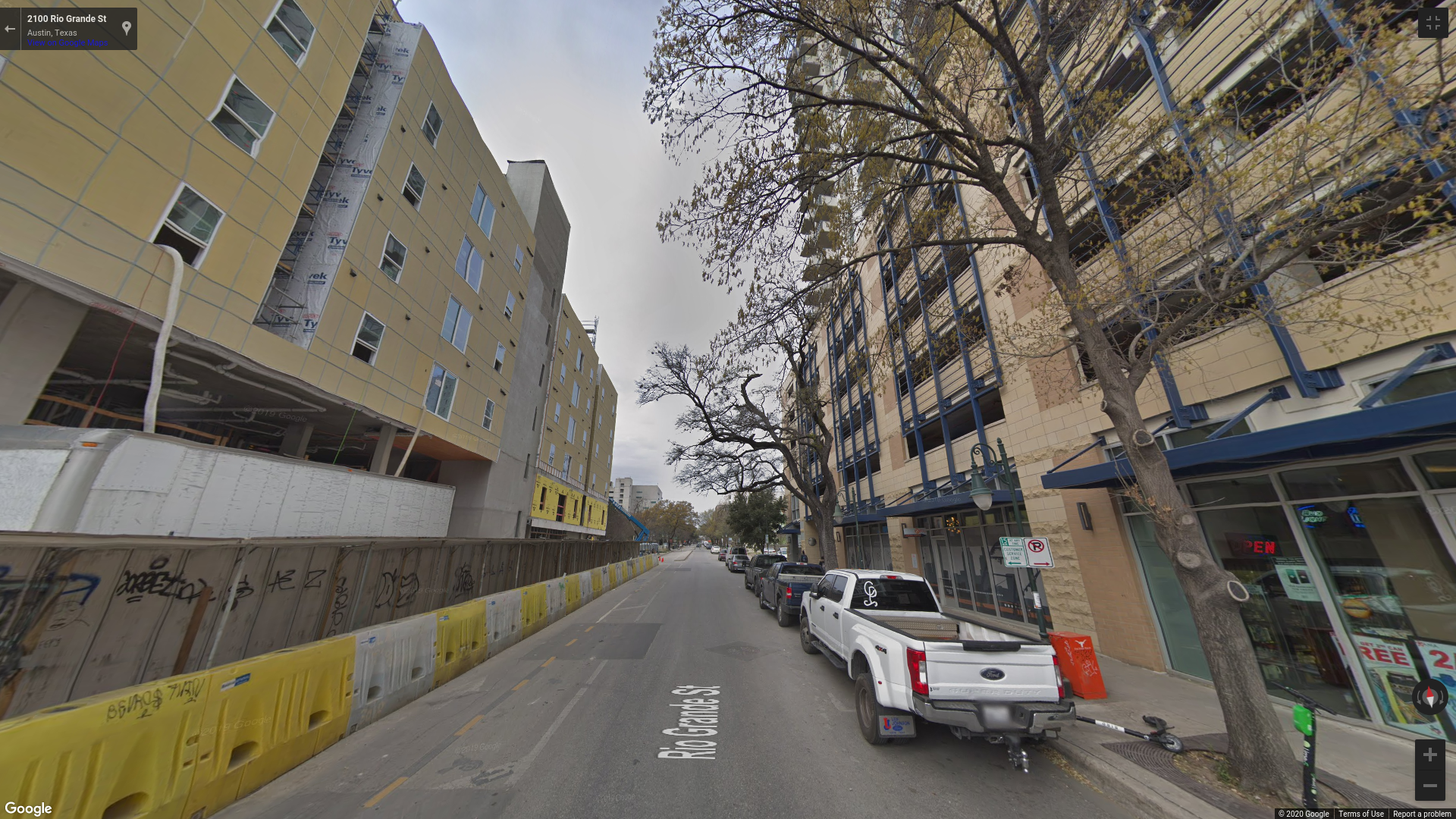}
	\end{minipage}
	\begin{minipage}[b]{0.325\textwidth}
		\centering
		\includegraphics[width=\linewidth]{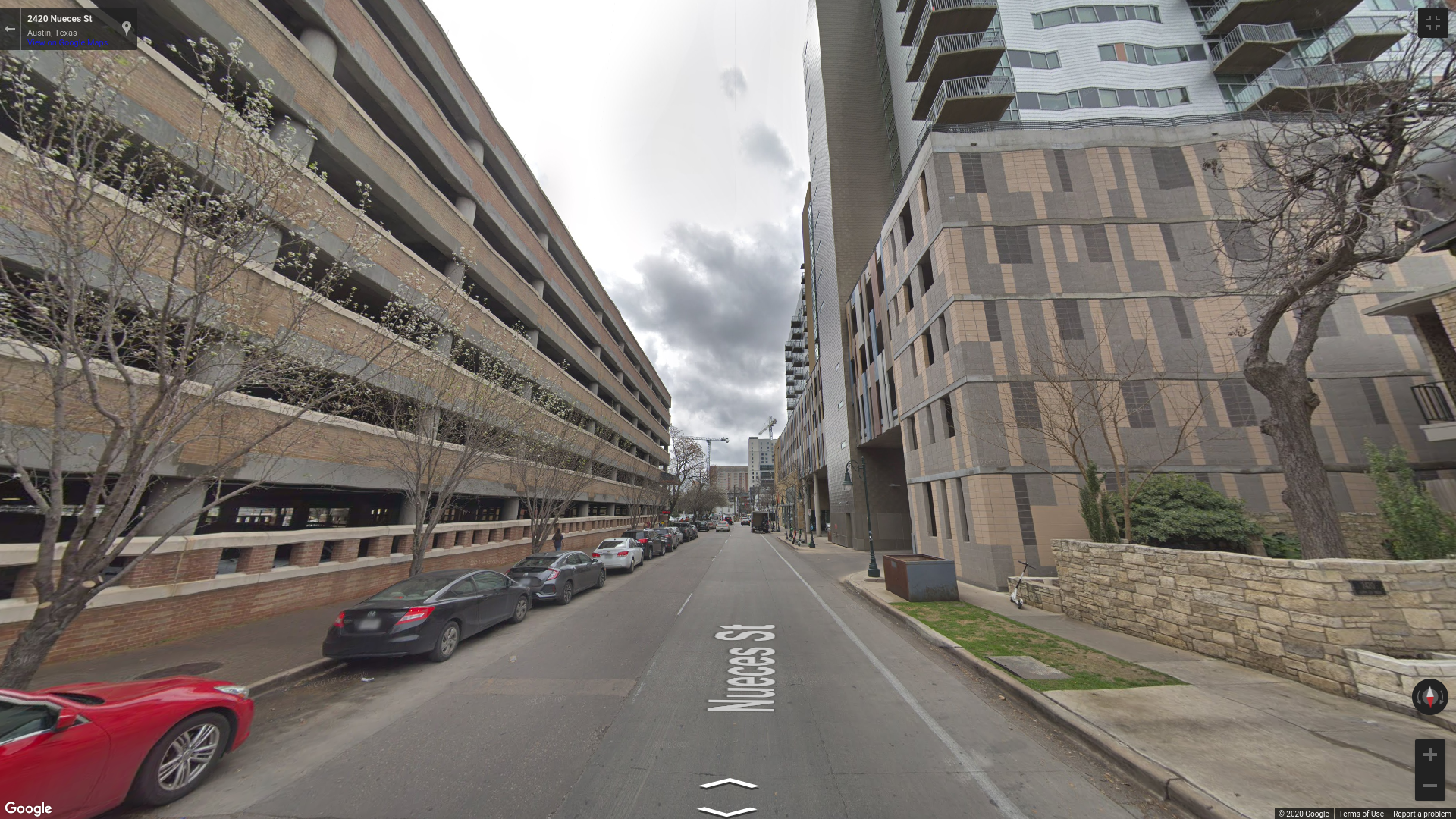}
	\end{minipage} \par\medskip
	\begin{minipage}[b]{0.325\textwidth}
		\centering
		\includegraphics[width=\linewidth]{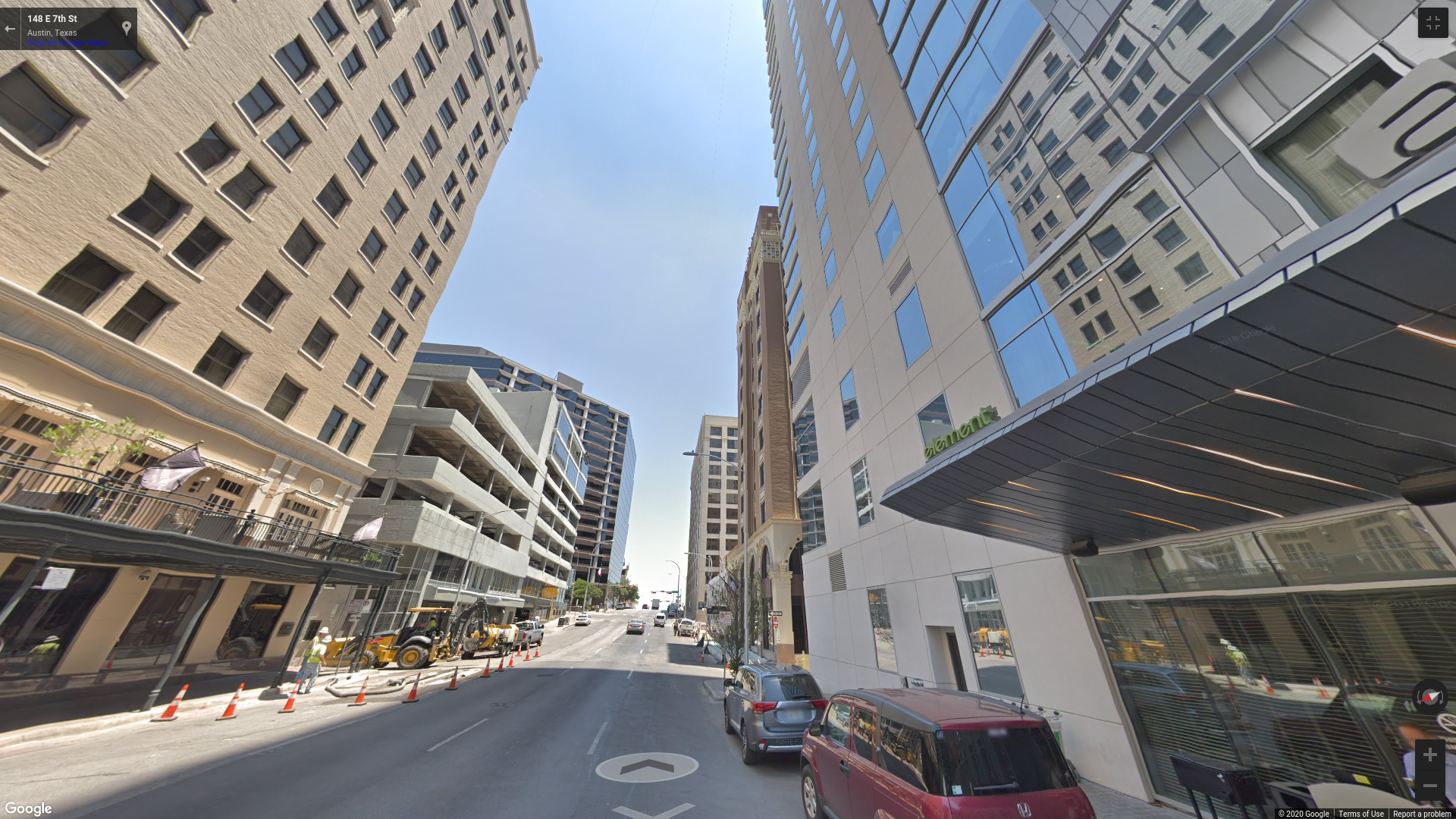}
	\end{minipage}
	\begin{minipage}[b]{0.325\textwidth}
		\centering
		\includegraphics[width=\linewidth]{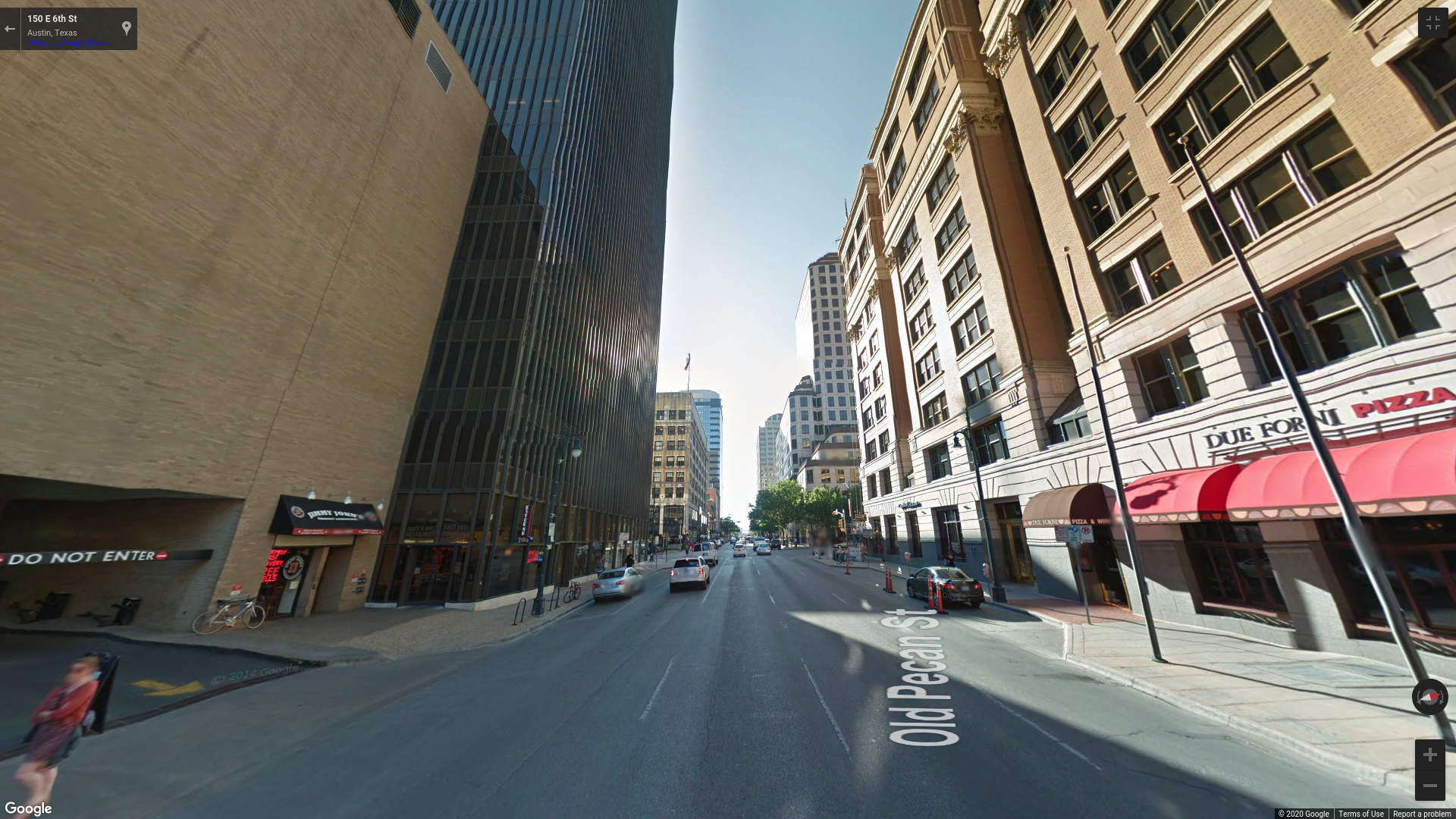}
	\end{minipage}
	\begin{minipage}[b]{0.325\textwidth}
		\centering
		\includegraphics[width=\linewidth]{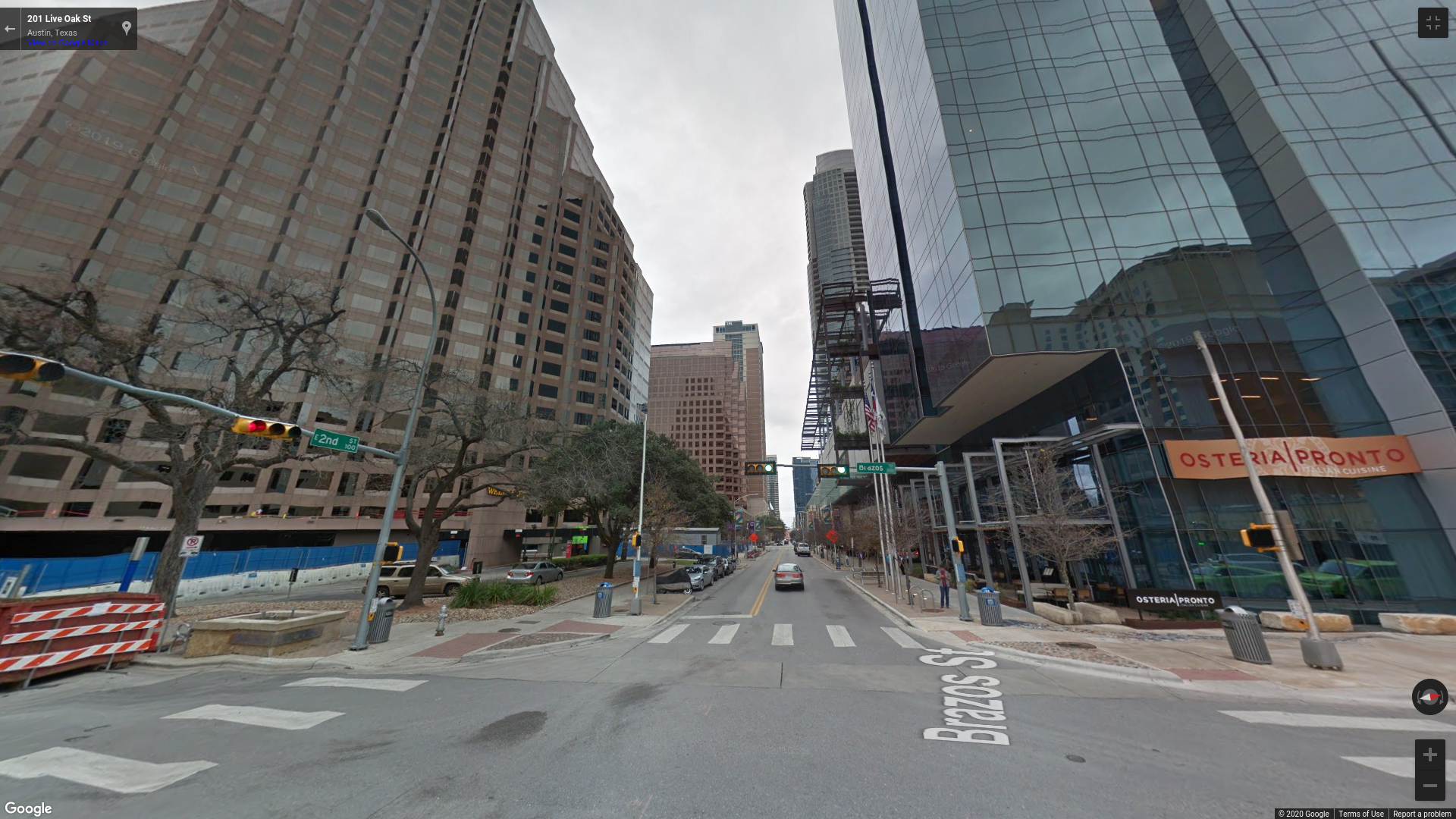}
	\end{minipage} \par\medskip
	\begin{minipage}[b]{0.325\textwidth}
		\centering
		\includegraphics[width=\linewidth]{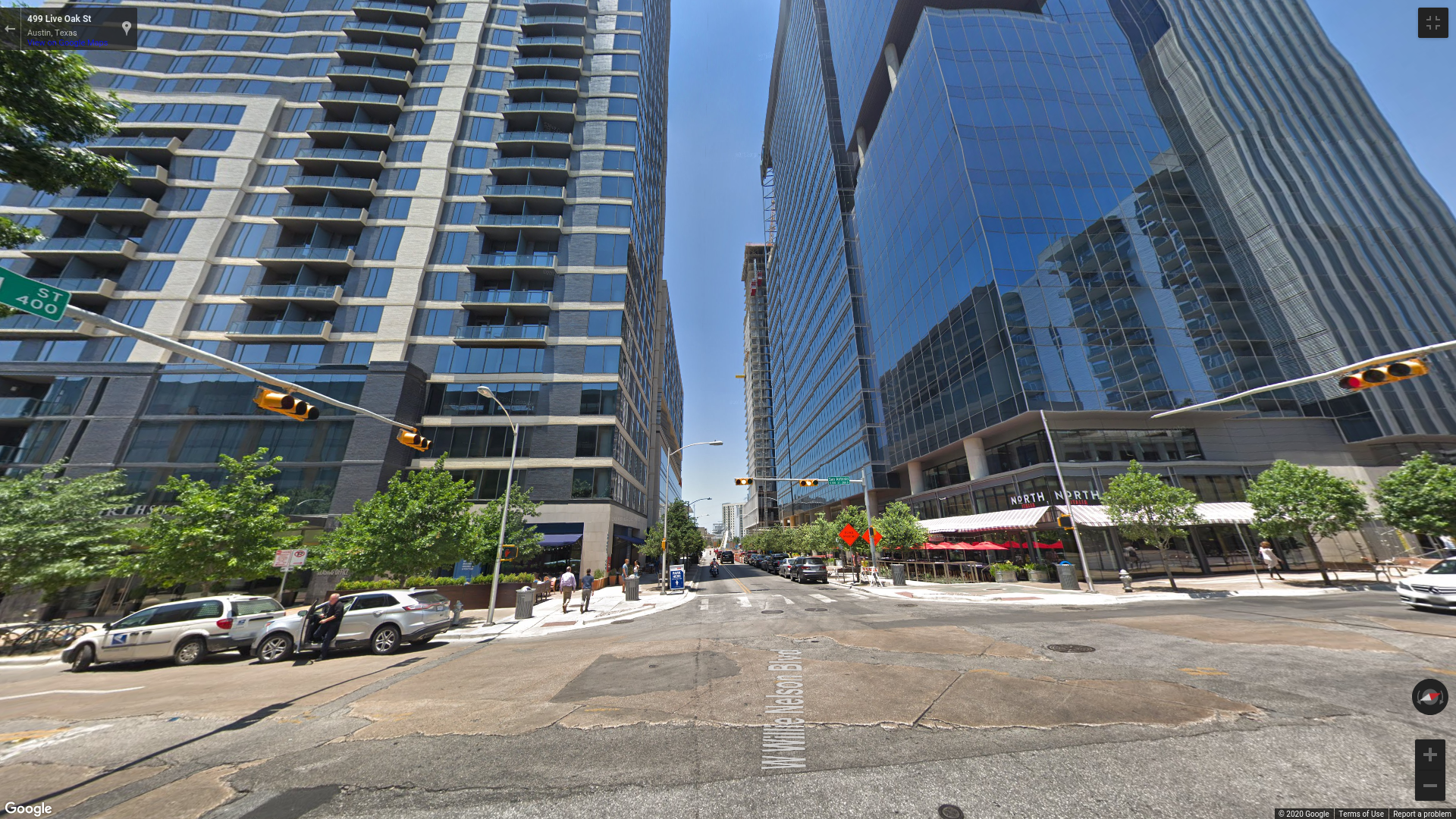}
	\end{minipage}
	\begin{minipage}[b]{0.325\textwidth}
		\centering
		\includegraphics[width=\linewidth]{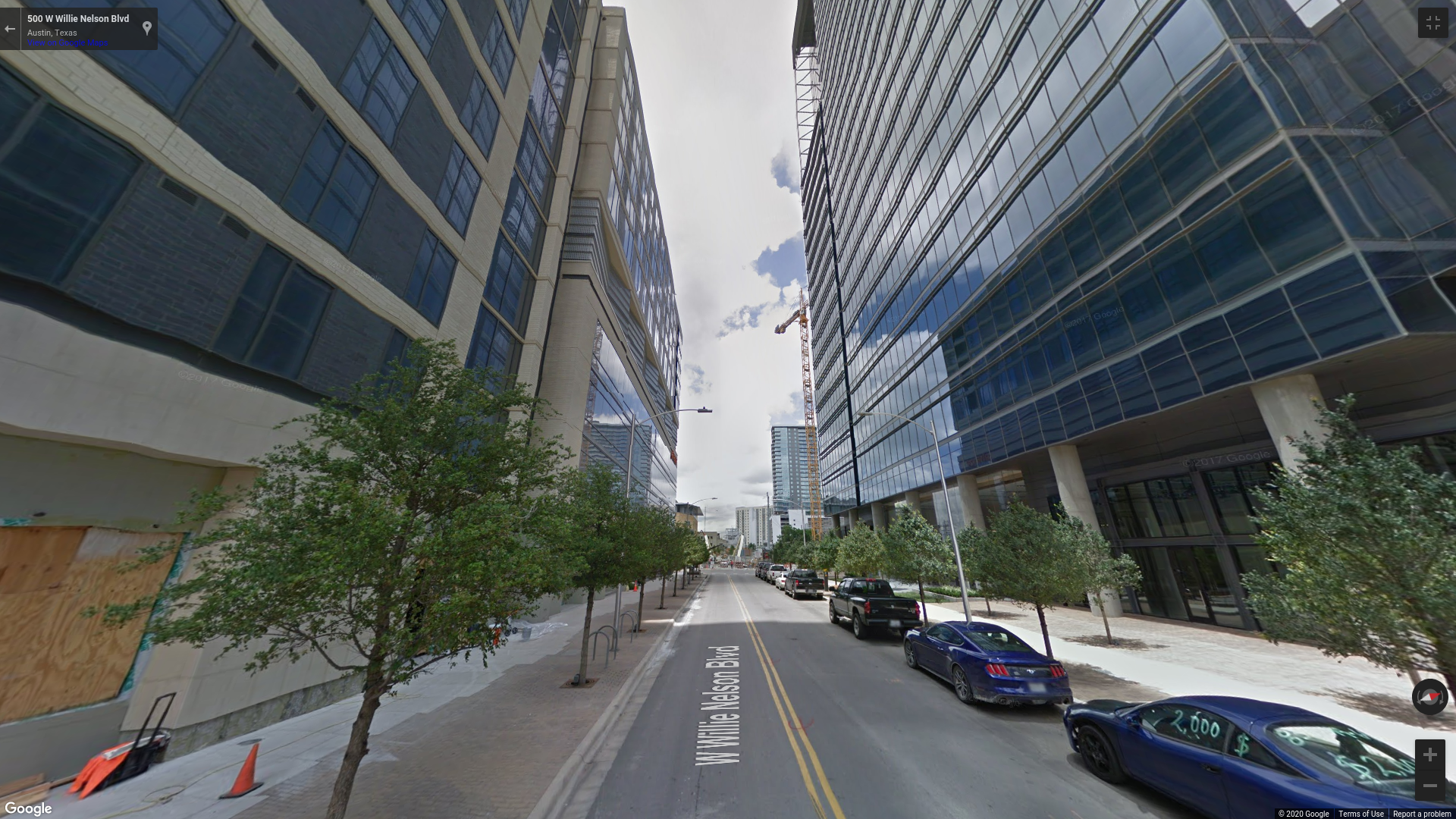}
	\end{minipage}
	\begin{minipage}[b]{0.325\textwidth}
		\centering
		\includegraphics[width=\linewidth]{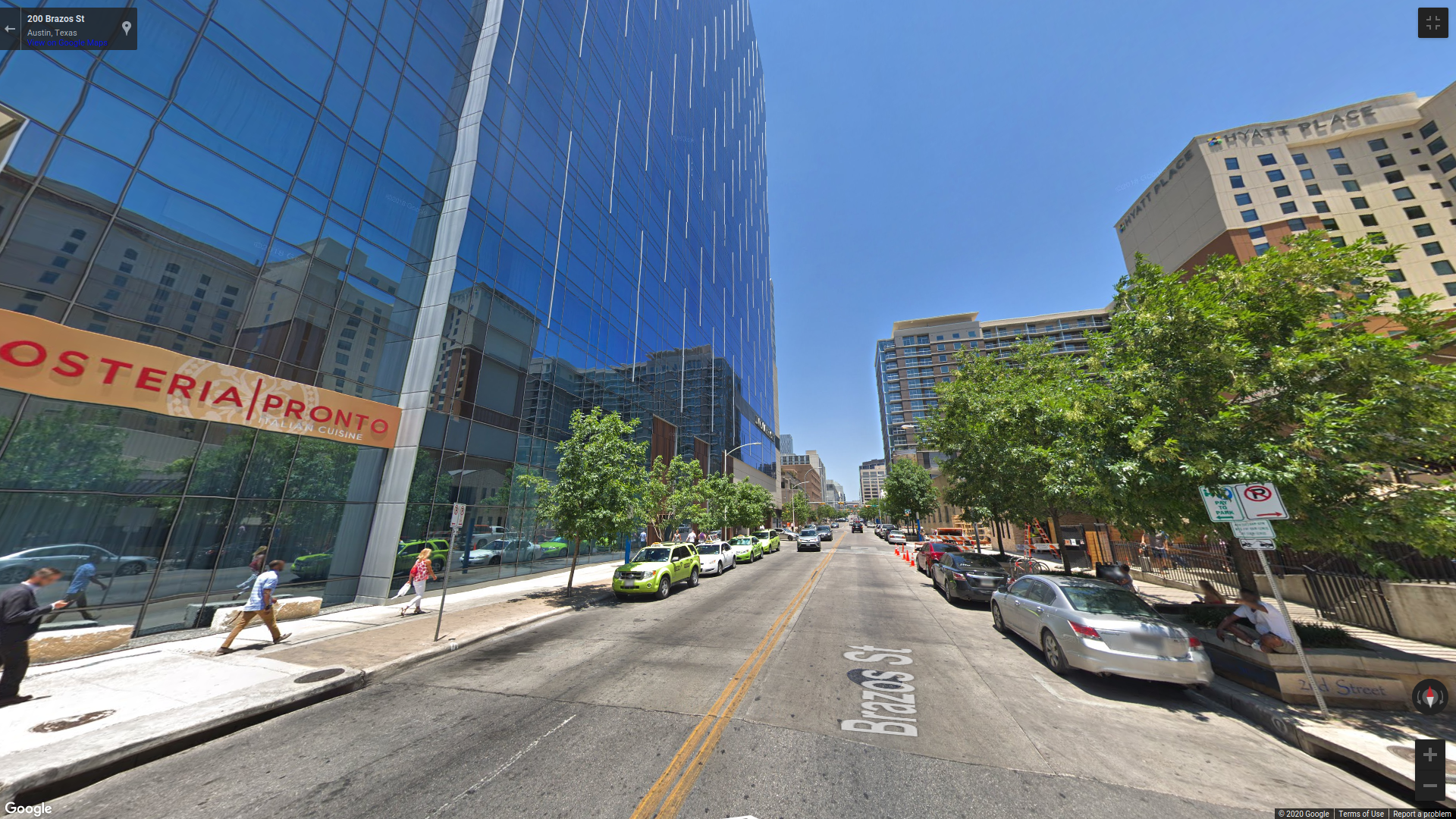}
	\end{minipage}
	\caption{Google Street View imagery of a few challenging scenarios
		encountered in the TEXCUP dataset \cite{narula2020texcup}.}
	\label{fig:atx-dataset}
\end{figure*}

Data was gathered on the UT Radionavigation Laboratory (RNL) \emph{Sensorium},
an integrated platform for automated and connected vehicle perception
research. It is equipped with multiple radars, IMUs, GNSS receivers, and a
lidar, as shown in Fig.  \ref{fig:sensorium}.  With the \emph{Sensorium}, the
RNL produced a public benchmark dataset collected in the dense urban center of
the city of Austin, TX called TEX-CUP \cite{narula2020texcup} for evaluating
multi-sensor GNSS-based urban positioning algorithms
\cite{humphreys2019deepUrbanIts}. The data captured includes a diverse set of
multipath environments (open-sky, shallow urban, and deep urban) as shown in
Fig.  \ref{fig:atx-dataset}. The TEX-CUP dataset provides raw wideband IF GNSS
data with tightly synchronized raw measurements from multiple IMUs and a
stereoscopic camera unit, as well as truth positioning data. This allows
researchers to develop algorithms using any subset of the sensor measurements
and compare their results with the true position.

For this paper's analysis, only the raw GNSS intermediate-frequency (IF)
samples from the primary antenna and inertial data from TEX-CUP were
considered, along with inertial data.  Two-bit-quantized IF samples were
captured at the Sensorium and at the reference station through the
\emph{RadioLynx}, a low-cost L1+L2 GNSS front end with a 5 MHz bandwidth at
each frequency, and were processed with the RNL's GRID SDR
\cite{t_humphreys06_scp,lightsey2013demonstration,t_humphreys09_sdgr,yoder2020visonFusion,clements2021bitpackingIonGnss}.
The tightly-coupled CDGNSS estimator described earlier was implemented in C++
as a new version of the GRID's sensor fusion engine.  The system's performance
was separately evaluated using inertial data from each of the Sensorium's two
MEMS inertial sensors. The first, a LORD MicroStrain 3DM-GX5-25, is an
industrial-grade sensor. The second, a Bosch BMX055, is a surface-mount
consumer-grade sensor.

TEX-CUP provides ground truth data for the vehicle position and orientation.
The truth dataset was generated by a combination of sensor fusion and a
tactical-grade IMU. The \emph{Sensorium} is equipped with an iXblue ATLANS-C: a
high-performance RTK-GNSS coupled fiber-optic gyroscope inertial navigation
system.  The post-processed fused RTK-INS position solution obtained from the
ATLANS-C is taken to be the ground truth trajectory.  Post-processing software
provided by iXblue generates a forward-backward smoothed position and
orientation solution with fusion of AsteRx4 RTK solutions and inertial
measurements. The post-processed solution is accurate to better than 10
centimeters throughout the dataset.  The effectiveness of the developed
spoofing detection method is evaluated with these datasets.

\section*{Spoofing Methodology}
The total signal at the victim receiver antenna is
\begin{align*}
  y_{\mr{tot}}(t) = y_{{a}}(t) + y_{{s}}(t) + \nu(t)
\end{align*}
where $y_{{a}}(t)$ is the authentic signal, $y_{{s}}(t)$ is the
spoofed signal, and $\nu(t)$ is the received noise.  Under a challenging
spoofing attack, $y_{{s}}(t)$ contains a perfect null of the authentic
signal and $\nu(t)$ is entirely naturally generated, i.e., not introduced by the
spoofer.  

\subsection*{Physical-Layer Spoofing} To artificially simulate a spoofing
attack, over-the-air, cable injection, and digital signal injection spoofing
were considered.  Over-the-air attacks are possible \cite{kerns2014unmanned,
  shepard2012droneHack, bhatti2017hostile}, but are not authorized in urban
areas.  A cable injection attack would be permissible for a live experiment in
an urban area, and digital signal combining, as in \cite{t_humphreys2014stb},
is a powerful after-the-fact spoofing technique.  But in both cases it is
challenging to explore a \emph{worst case} spoofing attack in which the
authentic signals $y_{\mr a}$ are entirely nulled by an antipodal spoofing
signal, as described in \cite{t_humphreys2014stb}.  Experience with {\tt ds7}
and {\tt ds8} from the Texas Spofing Test Battery (TEXBAT,
\cite{humphreys2012_TEST_Battery,rnlTexbatSite}) revealed that such antipodal
spoofing is difficult to maintain under even static laboratory conditions.  The
remnant authentic signal from an unsophisticated and imperfect spoofing attack
sullies the test statistic, making detection too easy and leading to an overly
optimistic performance assessment.

Of course, nulling $y_{\rm a}$ can also be achieved by generating a spoofing
signal so powerful that it buries the authentic signal below the receiver's
noise floor.  But this should not be considered a \emph{worst case} attack
because the overwhelmingly high received signal power from the spoofer is
easily detected \cite{wesson2018pincer}.

In short, physical-layer spoofing is challenging to conduct in such a way as to
present a convincing worst-case spoofing attack to this paper's detector.

\subsection*{Observation-Domain Spoofing}
It is important to evaluate spoofing detection techniques on a worst-case
spoofing attack, with the idea being that if the proposed detection strategy is
effective on the worst-case scenario, it is even more effective on weaker
attacks.  Accordingly, this paper adopts \emph{observation-domain spoofing}.
The spoofing in the observation domain is advantageous because the authentic
signal is inherently nulled, presenting a subtle attack.

The first method of implementing observation-domain spoofing is \emph{position
offset spoofing}.  With position offset spoofing, a position offset $\delta \vb
r$ is added to the authentic measured position $\delta \vb r_a$ to generate a
spoofed position $\vb r_s = \vb r_a + \delta \vb r(t)$.  This is accomplished
by altering the pseudorange and carrier phase measurements from each satellite
so that they correspond to the spoofed position with the desired additive
position offset $\delta \vb r$.  The spoofed pseudorange $\rho^i_s(t)$ and
carrier phase $\phi^i_s(t)$ measurements for the $i$th satellite are
constructed as follows 
\begin{equation}
	\label{eq:pr_ps}
	\rho^i_s(t) = \rho^i_a(t) + \delta \rho^i(\delta \vb r(t))
\end{equation}
\begin{equation}
	\label{eq:cp_ps}
	\phi^i_s(t) = \phi^i_a(t) + \delta \phi^i(\delta \vb r(t))
\end{equation}
where $\rho^i_a(t)$ and $\phi^i_a(t)$ are the authentic pseudorange and carrier
phase for the $i$th satellite and $\delta \rho^i(\cdot)$ and
$\delta \phi^i(\cdot)$ are the nonlinear (but easily linearizable) mapping
functions, based on the geometry for the particular $i$th satellite, that map
the position offset $\delta \vb r(t)$ to the corresponding pseudorange and
carrier phase offsets.

The second method of implementing observation-domain spoofing is
\emph{timestamp spoofing}.  With timestamp spoofing, the measurements at a
particular time are reassigned to have an alternate measurement timestamp.  The
required modifications to the spoofed pseudorange $\rho^i_s(t)$ and carrier
phase $\phi^i_s(t)$ measurement from each satellite induced by the new
timestamp may be expressed as 
\begin{equation}
	\label{eq:pr_ts}
	\rho^i_s(t) = \rho^i_a(t + \delta t(t)) + \Delta \rho^i(t, \delta t(t))
\end{equation}
\begin{equation}
	\label{eq:cp_ts}
	\phi^i_s(t) = \phi^i_a(t + \delta t(t)) + \Delta \phi^i(t, \delta t(t))
\end{equation}
where $\delta t(t)$ is the timestamp shift applied.  The authentic observables
from time $t + \delta t(t)$ are fed to the estimator as if they had occurred at
time $t$.  The functions $\Delta \rho^i(t, \delta t(t))$ and
$\Delta \phi^i(t, \delta t(t))$ adjust the timestamp-shifted observables to
account for the transmitting spacecraft's orbital motion and clock evolution
over the interval from $t$ to $t + \delta t(t)$.

In position offset spoofing, because
$\phi_s(t) = \phi_a(t) + \delta \phi(\delta \vb r(t))$, all vehicle motion
reflected in $\phi_a(t)$ is also present in $\phi_s(t)$.  This includes all
high-frequency motion due to the road irregularities and other minor movements.
A detection technique designed to detect small-amplitude, high-frequency
discrepancies in $\phi(t)$ via the WFARC would not actually see such
discrepancies unless $\delta \phi (\delta \vb r (t))$ also included simulated
high-frequency content.  

By contrast, timestamp spoofing borrows spoofed phase and pseudorange
measurements from a different time instant, ensuring that high-frequency
variations in these quantities will be different from those predicted by the
\emph{a priori} state based on IMU propagation.  This is more representative of
an actual spoofing attack scenario in which the attacker cannot predict the
high-frequency vehicle motion.  Moreover, by reducing the timestamp shift
$\delta t(t)$, one can realize ever-subtler attacks that are increasingly hard
to detect, allowing exploration of worst-case-for-detectability spoofing.

\begin{figure}[H]
  \centering
  \includegraphics[scale=.15]{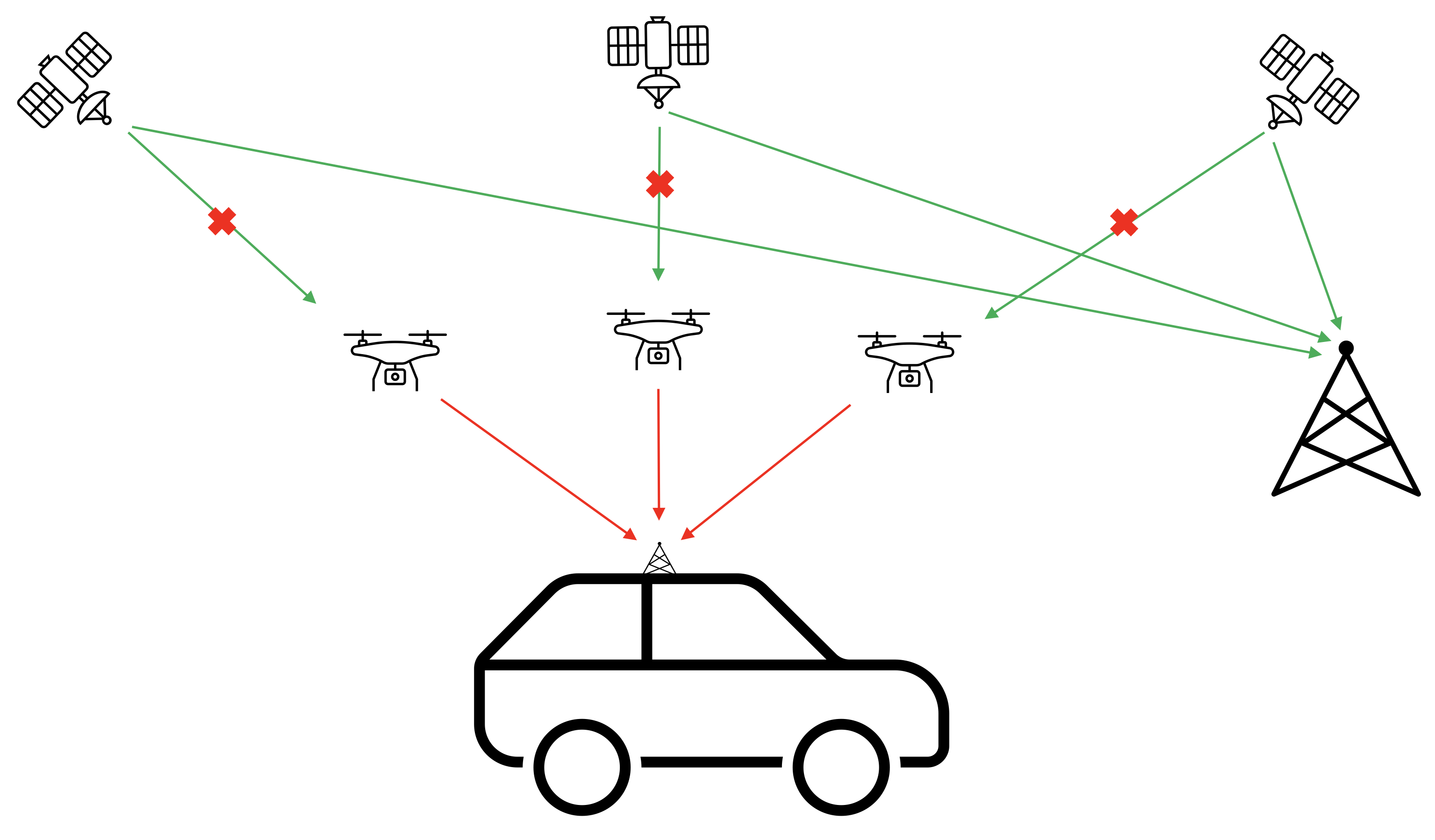}
  \caption{Timestamp spoofing is representative of a worst-case spoofing attack
    in which the attacker positions a fleet of drone-borne spoofers such that
    spoofing signals (1) perfectly null the corresponding authentic signals,
    and (2) emanate from positions along the line of sight connecting each
    overhead GNSS satellite to the target vehicle.  The reference receiver
    shown to the right, whose GNSS observables are used for precise CDGNSS
    processing, is presumed to be unaffected by the spoofing attack.}
  \label{fig:attack}
\end{figure}

Timestamp spoofing is also worst-case in a different sense.  Because each
spoofed observable is made to be consistent with the GNSS constellation
geometry, the spoofing detector is presented with observables whose implied
geometry is consistent with actual transmitter locations, not, for example,
with a single transmitting spoofer.  Moreover, because there are no
irregularities in the observables that might result from an over-the-air or
direct-injection attack in which the authentic signals are not perfectly nulled
and so conflict with the spoofing signals, the spoofing detector is presented
with observables whose variations, both in amplitude and frequency content, are
entirely plausible.  Thus, timestamp spoofing is representative of a case in
which a well-financed attacker is able to place a
single-satellite-full-single-ensemble spoofer capable of full authentic-signal
nulling along the line-of-sight from the target vehicle to each overhead GNSS
satellite, as illustrated in Fig.  \ref{fig:attack}.

\section*{Results}

The following section presents an analysis of the proposed test statistic in
both the non-spoofing case and against a worst-case attack.  Results with the
industrial- and consumer-grade IMU are presented.

\subsection*{Characterization of the Null Hypothesis}
This spoofing detector is premised on a hypothesis test between statistical
models for the authentic and counterfeit GNSS signals.  The statistics of the
null hypothesis must be fully characterized so that a statistical baseline is
established, against which carrier phase errors induced by spoofing in the same
setting can be compared.  The null hypothesis of dynamic ground vehicle
scenarios includes natural effects such as blockage and multipath, which is the
predominant source of error.

To analyze the null hypothesis, the WFARC was calculated
in the nominal case through the entirety of the TEX-CUP dataset
containing no spoofing.  Because multipath is dependent on the surrounding
environment, two categories were separately considered: shallow urban and deep
urban.  Measurements were separated into these categories manually by
identifying segments of the dataset where the vehicle resided in shallow urban
and deep urban areas.

\begin{figure}[H]
	\centering
	\begin{minipage}{.5\textwidth}
		\centering
		\includegraphics[width=1\linewidth]{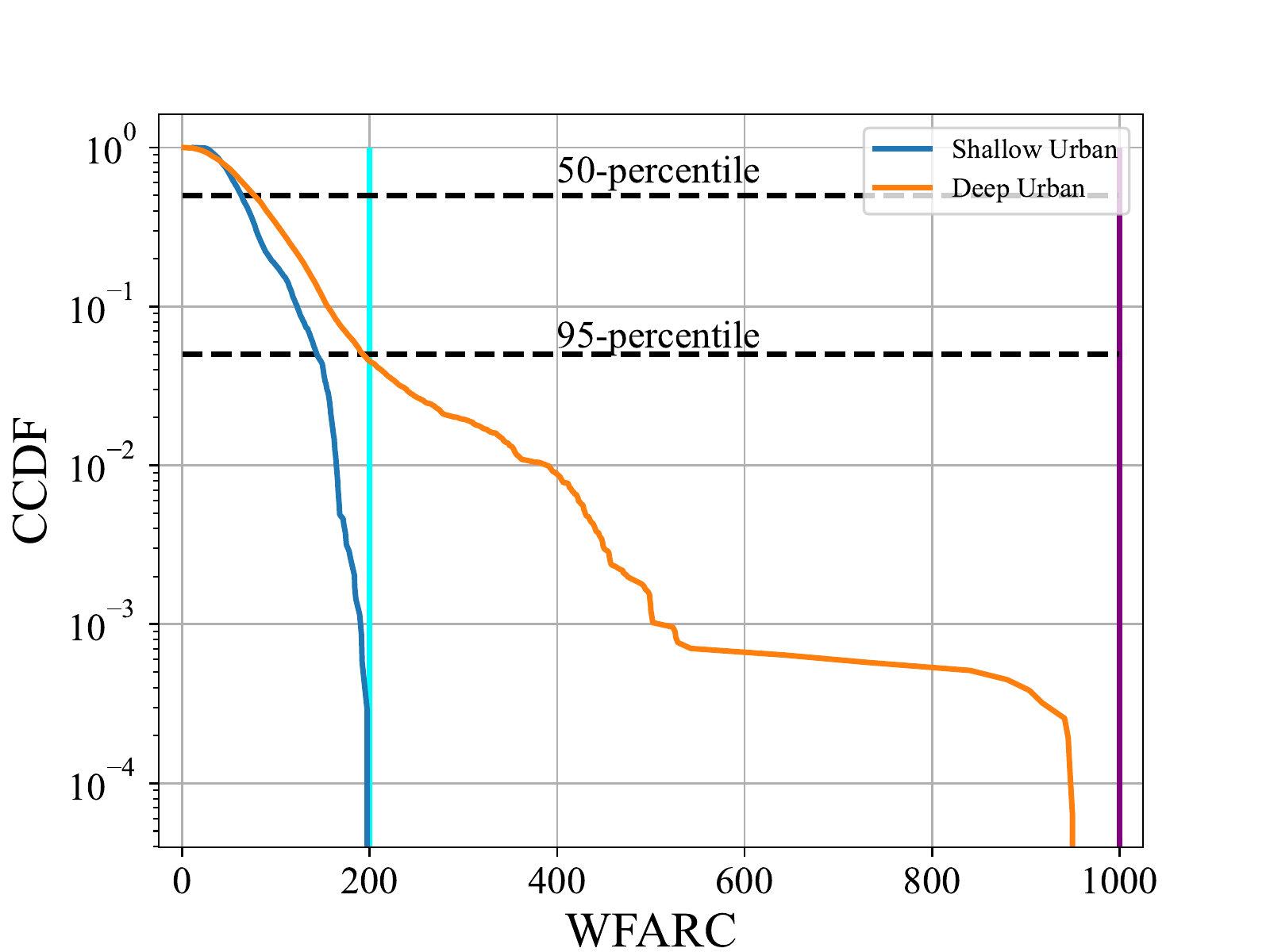}
	\end{minipage}%
	\begin{minipage}{.5\textwidth}
		\centering
		\includegraphics[width=1\linewidth]{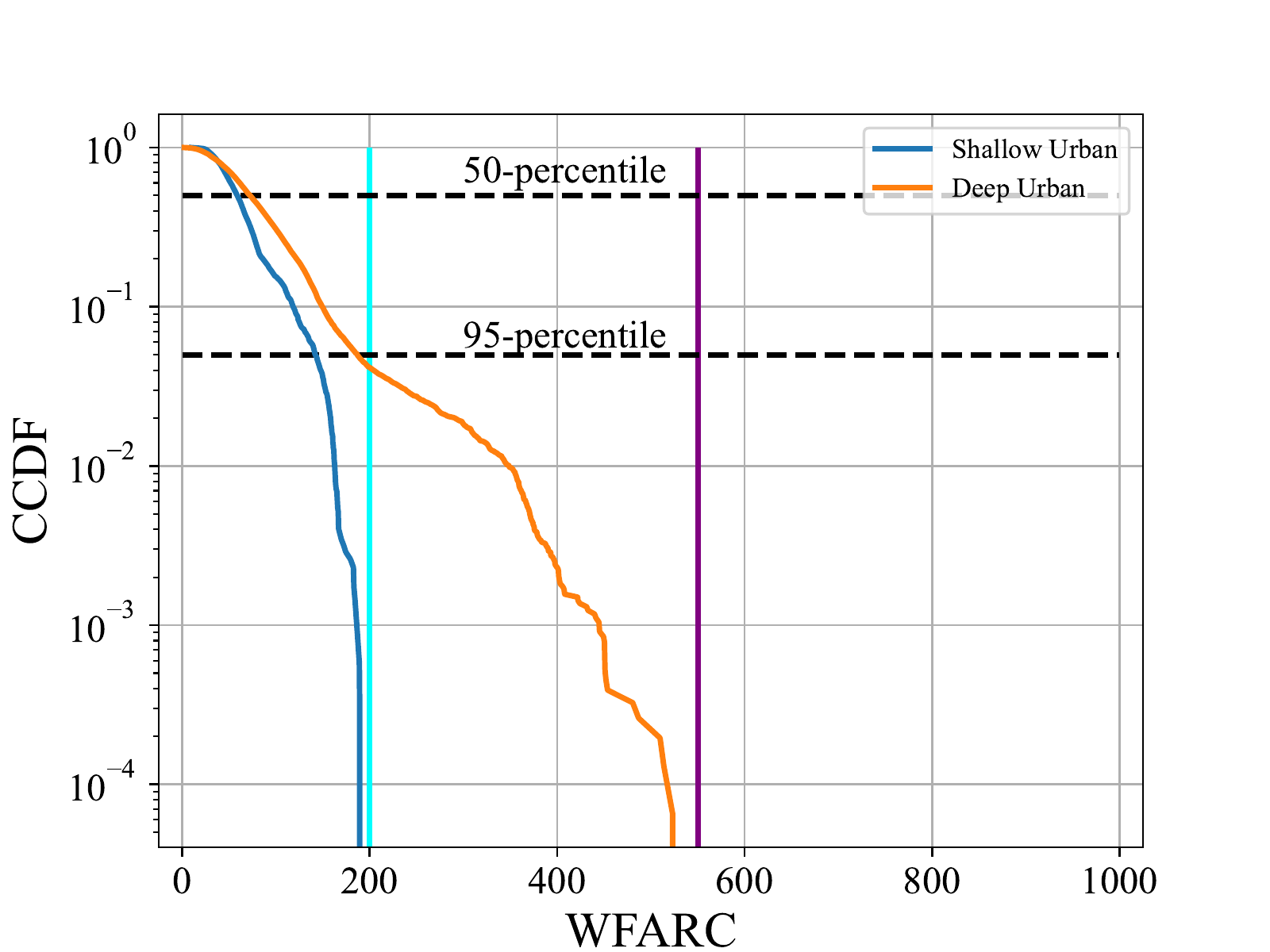}
	\end{minipage}
	\caption{The complementary cumulative distribution function (CCDF) of the WFARC over the entire TEX-CUP dataset with the LORD MicroStrain 3DM-GX5-25 (industrial grade) IMU on the left and with the Bosch BMX055 (consumer grade) IMU on the right. The test statistic is separated into two categories: shallow urban and deep urban.  The cyan line represents the largest value of the WFARC in the shallow urban environment and the purple line represents the largest value of the WFARC in the deep urban environment.  The deep urban environment has a significantly longer tail compared ot the shallow urban environment.}
	\label{fig:H0_hist}
\end{figure}

\begin{figure}[H]
	\centering
	\begin{minipage}{.5\textwidth}
		\centering
		\includegraphics[width=1\linewidth]{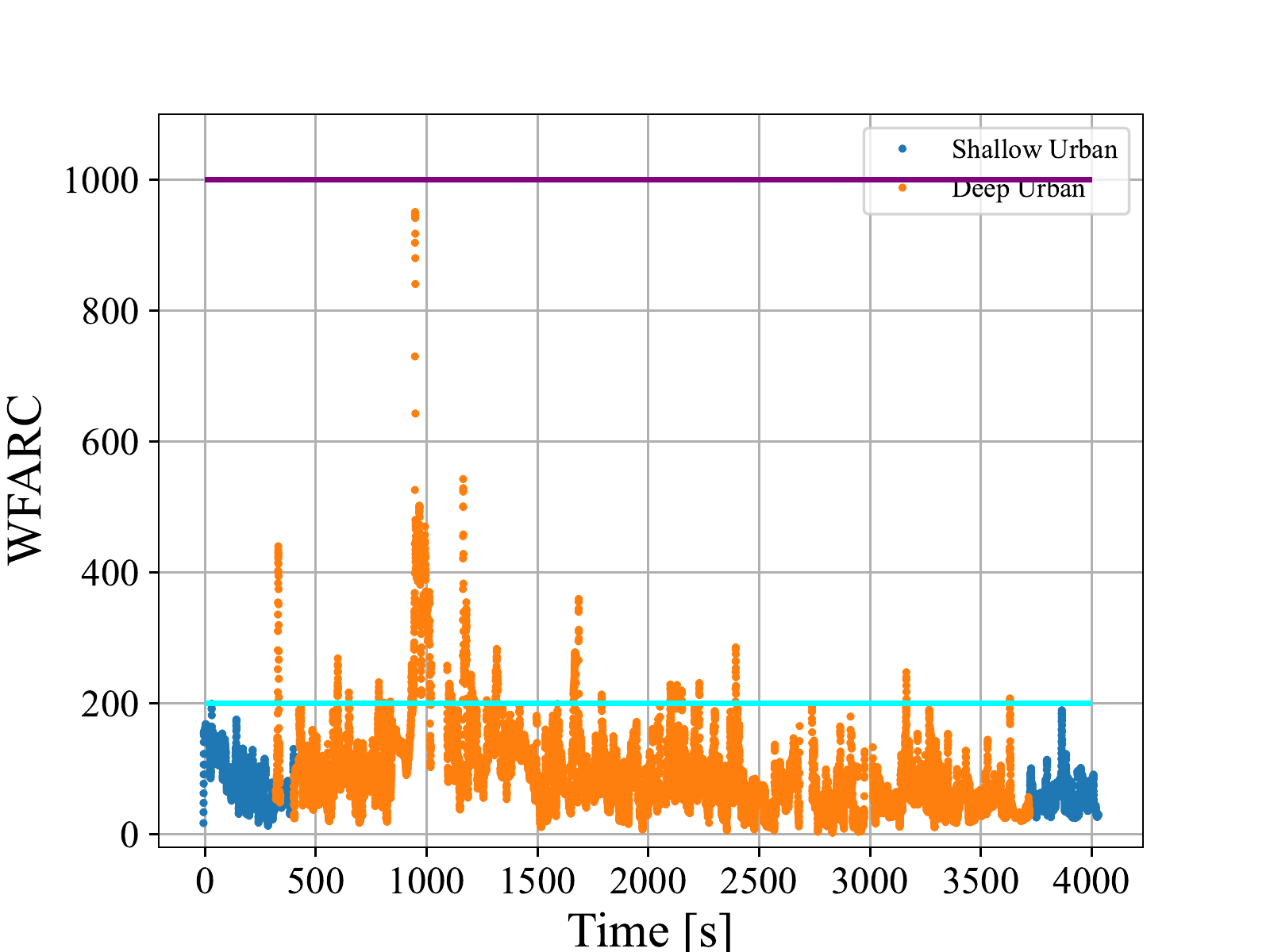}
	\end{minipage}%
	\begin{minipage}{.5\textwidth}
		\centering
		\includegraphics[width=1\linewidth]{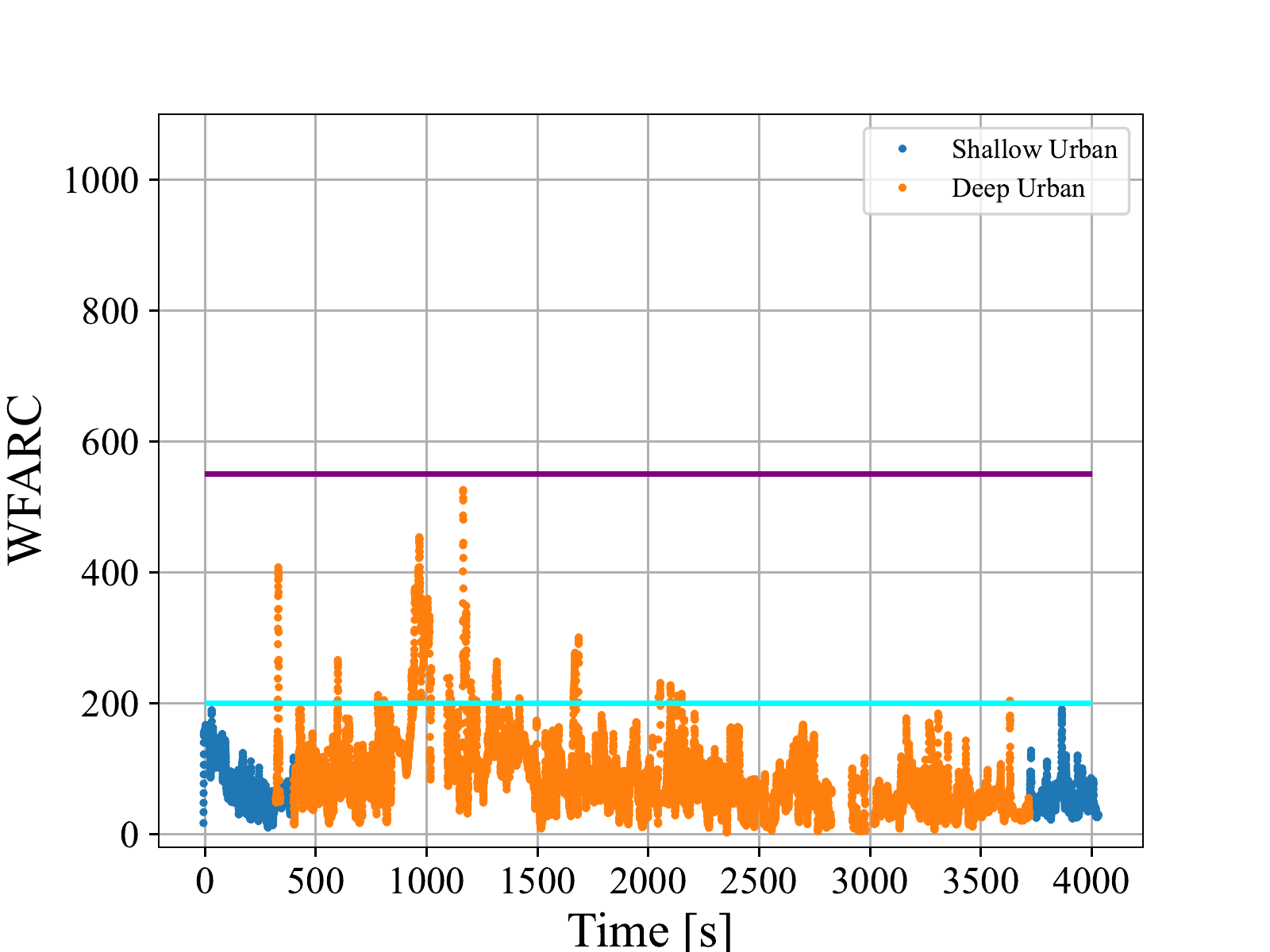}
	\end{minipage}
	\caption{A time history of the WFARC over the entire TEX-CUP dataset with the industrial grade IMU on the left and the consumer grade IMU on the right. }
	\label{fig:H0_timehist}
\end{figure}

Fig. \ref{fig:H0_hist} shows the complementary cumulative distribution function
(CCDF) of the WFARC in shallow and deep urban environments
for the nominal case with industrial- and consumer-grade IMUs.  The test
statistic in the deep urban case has a much longer tail, which is expected
because of the extreme multipath and blockage in deep urban areas.  The cyan
line represents the largest value of the WFARC in the
shallow urban environment and the purple line represents the largest value of
the WFARC in the deep urban environment.  These will be
the thresholds used to detect spoofing.  Because the test statistic in the null
hypothesis is never larger than these values, it corresponds to having a false
alarm probability of zero.  A chi-squared test can be used to lower these
thresholds but comes at the cost of having a fixed false positive rate.
Fig. \ref{fig:H0_timehist} shows the time history of the WFARC over the TEX-CUP dataset.

It is important to note that the WFARC while using the consumer grade IMU is
generally smaller than the WFARC while using the industrial grade IMU.  This is
expected because the consumer grade IMU is of lesser quality, thus having more
variance with each measurement.  The \emph{a priori} state estimate from IMU
tight coupling has a larger uncertainty because the estimator has less
confidence in the IMU measurements, leading corrupted measurements to be more
believable.  Once again, in the null hypothesis, spikes in the WFARC are caused
by multipath and blockage.

\subsection*{Performance Against a Worst-Case Spoofing Attack}

\begin{figure}[H]
	\centering
	\begin{minipage}{.5\textwidth}
		\centering
		\includegraphics[width=1\linewidth]{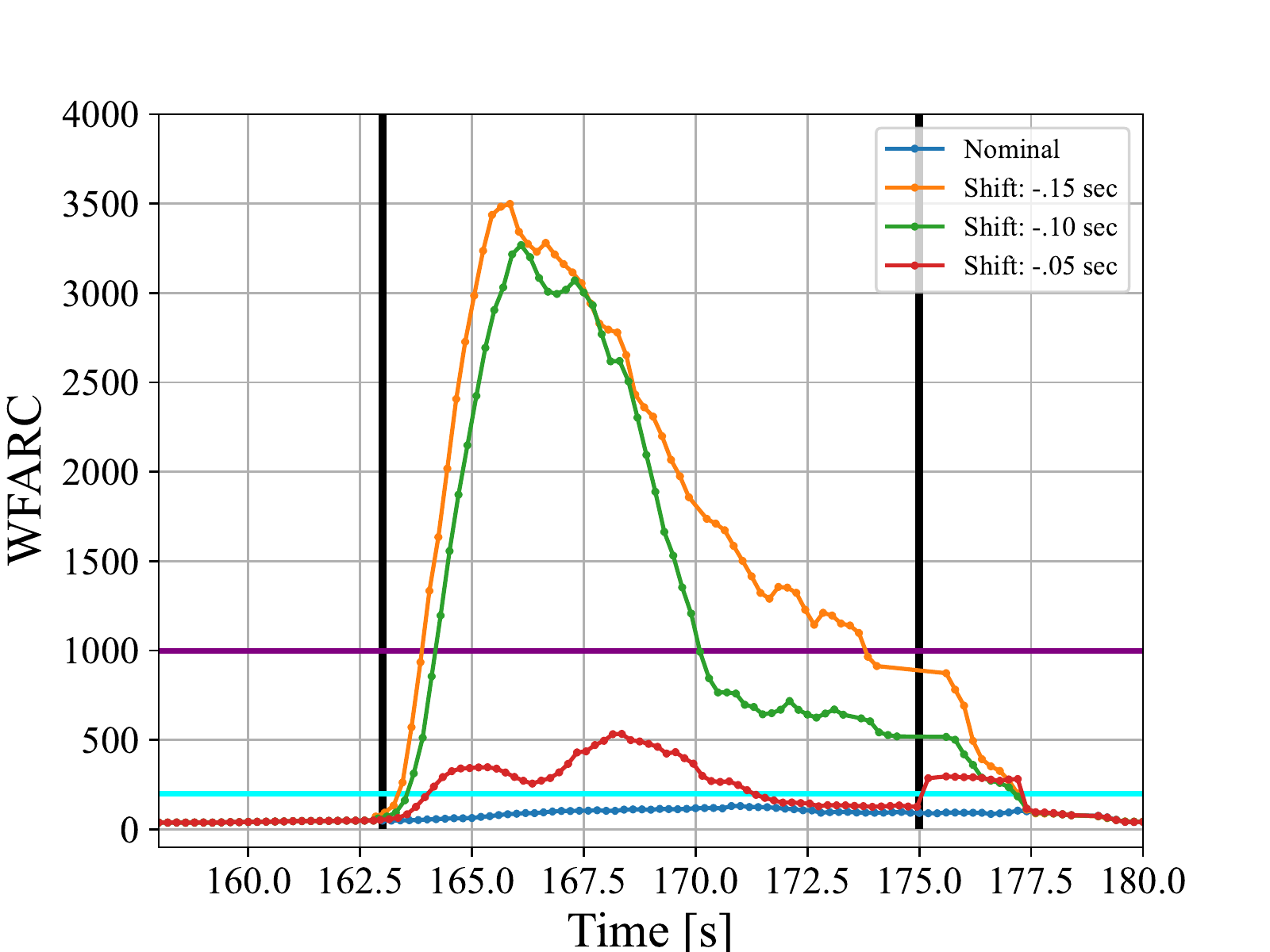}
	\end{minipage}%
	\begin{minipage}{.5\textwidth}
		\centering
		\includegraphics[width=1\linewidth]{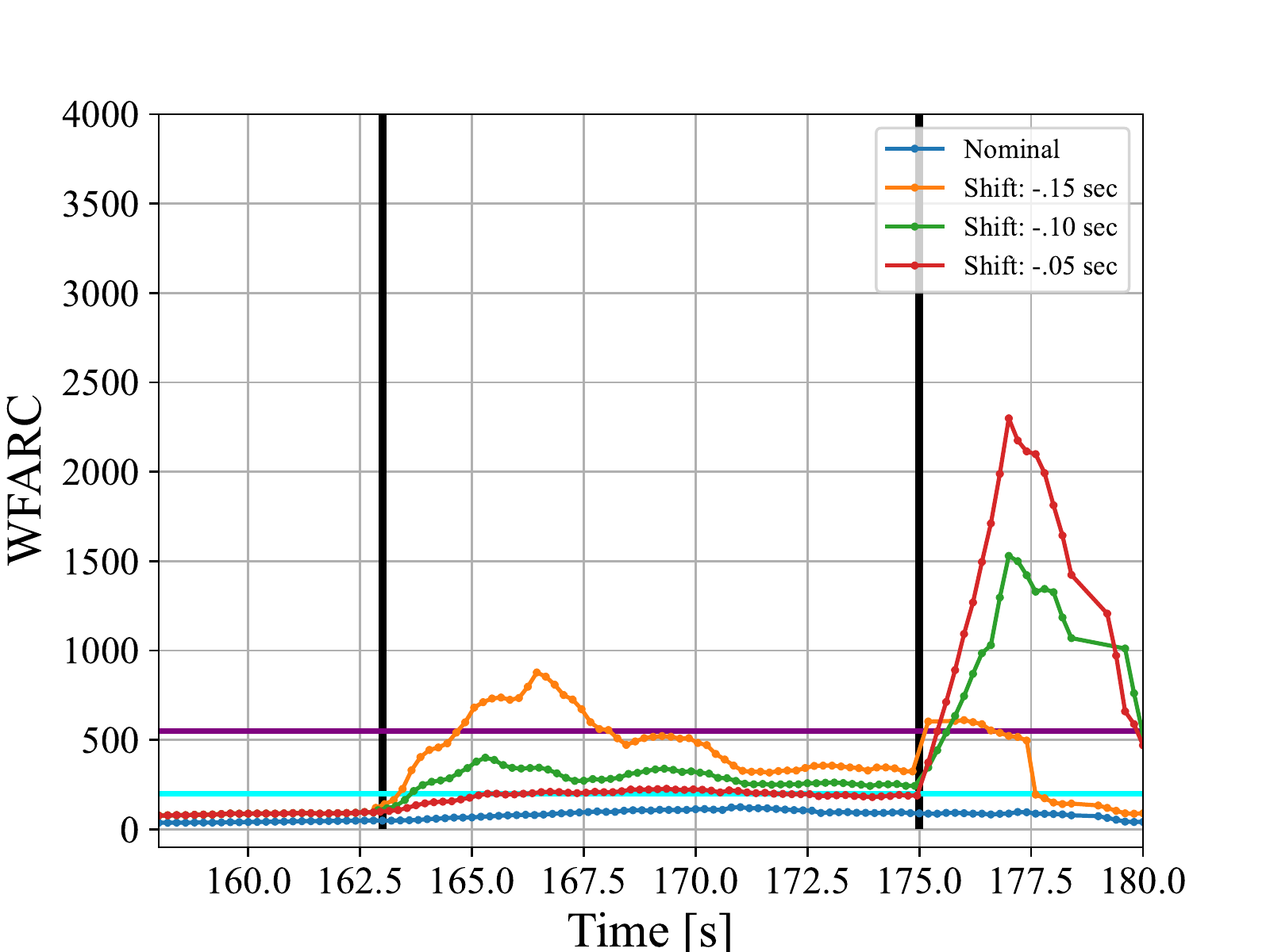}
	\end{minipage}
	\caption{WFARC during a worst-case spoofing attack in the shallow urban environment.  The plot on the left is with the LORD MicroStrain 3DM-GX5-25 (industrial grade) IMU and the plot on the right is with the Bosch BMX055 (consumer grade) IMU.  Spoofing begins at the 163 second mark and ends at the 175 second mark.  The vehicle is stopped at a red light, but then starts moving at the 163 second mark, introducing a small drag off from the true position. Plotted here are 3 different time shift attacks in the same scenario. The shift of .15 seconds is the least subtle attack while the .05 second attack is the most challenging attack.  From the analysis if the null hypothesis, the cyan line denotes the shallow urban threshold and the purple line denotes the deep urban threshold.}
	\label{fig:WinCPInno}
\end{figure}

The following is an example of a worst-case spoofing attack in a shallow urban
environment.  In this scenario, the spoofing attack begins while the vehicle is
stopped at a stoplight and continues as the vehicle begins to move.  
The WFARC in this scenario are shown in Fig. 
\ref{fig:WinCPInno} with both industrial- and consumer-grade IMUs.  
The vehicle starts moving at the 163 second mark.  The 
spoofing attack begins at the 163 second mark just before first movement 
and ends at the 175 second mark.  As the vehicle begins to
move, the position errors will grow gradually because the vehicle slowly begins
to accelerate forward, inducing a position error.  Three different time shift attacks 
in the same scenario are shown in this figure. The shift of .15 seconds is the 
least subtle attack while the .05 second attack is the most challenging attack 
because the faults are much smaller.  As the vehicle begins to move, the 
estimator recognizes inconsistencies between the spoofed GNSS measurements 
and the IMU because of the tight coupling.  The rise in the WFARC above the 
thresholds shows this disagreement that is attributed to spoofing.  

With the industrial grade IMU and using the shallow urban threshold, all three
time shifts spoofing attacks were identified within a second.  The estimator knows 
that the IMU data are different than the GNSS measurements from the WFARC,
much more than anything multipath would induce in the shallow urban environment.
If the vehicle was in the deep urban environment, the .05 second shift spoofing
attack would just be attributed to multipath.  The sensitivity of the test is
dependent on multipath environment. 

All three attacks were identified while using the consumer-grade IMU within two
seconds. If the deep urban threshold was applied, only the least challenging
attack would have been identified.  In all cases, the  WFARC is significantly
smaller compared to when the industrial IMU is used.  Once again, this is
because the estimator has more confidence in the spoofed GNSS measurements than
the lower quality IMU.  Interestingly, there is a spike in the WFARC after the
spoofing attack is over.  This happens because the estimator is showing trauma
from the spoofing attack-- the abrupt return of the true GNSS measurements were
significantly different from what the previously ingested spoofed measurements
were predicting.

\begin{figure}[H]
	\centering
	\begin{minipage}{.5\textwidth}
		\centering
		\includegraphics[width=1\linewidth]{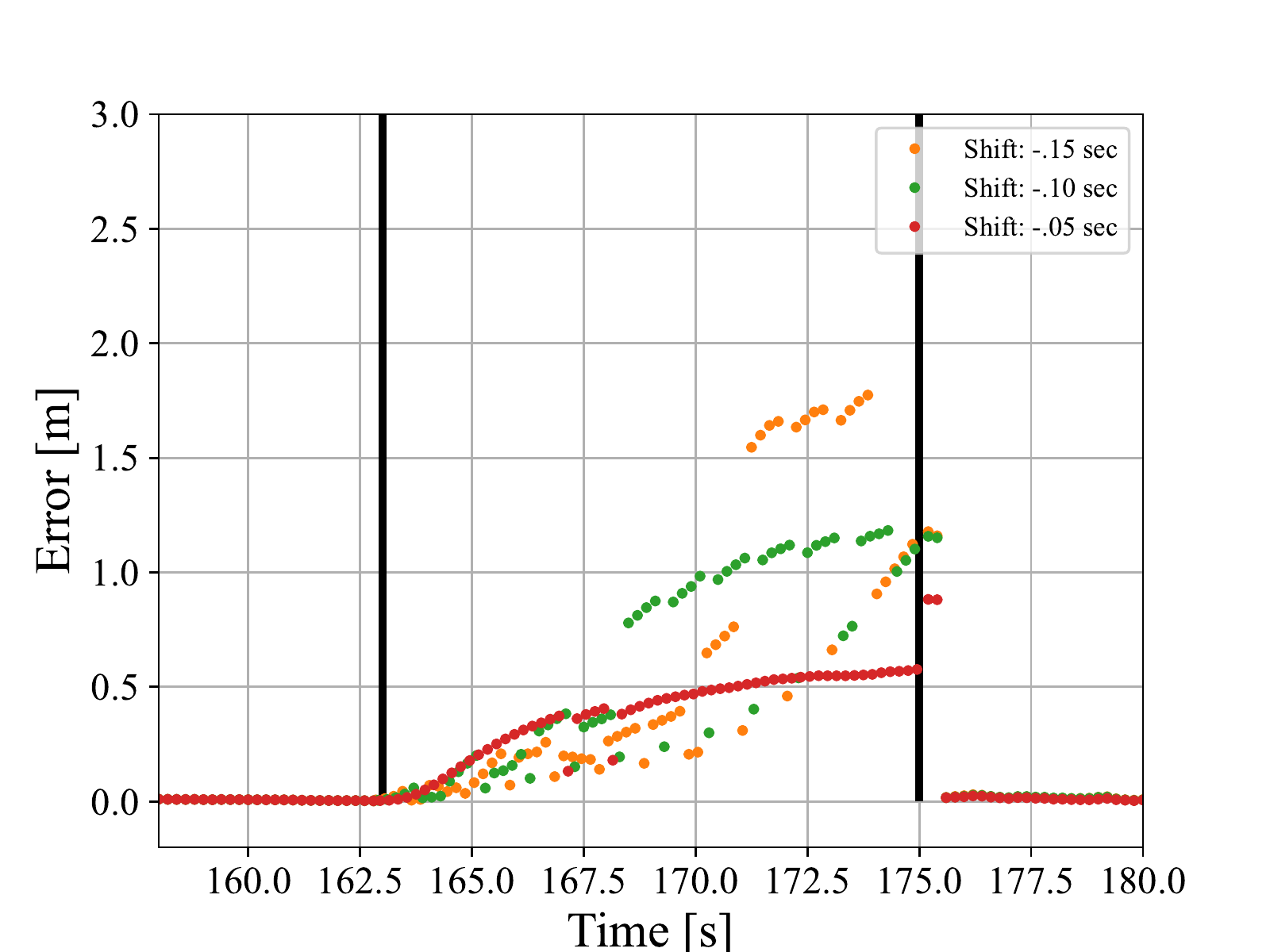}
	\end{minipage}%
	\begin{minipage}{.5\textwidth}
		\centering
		\includegraphics[width=1\linewidth]{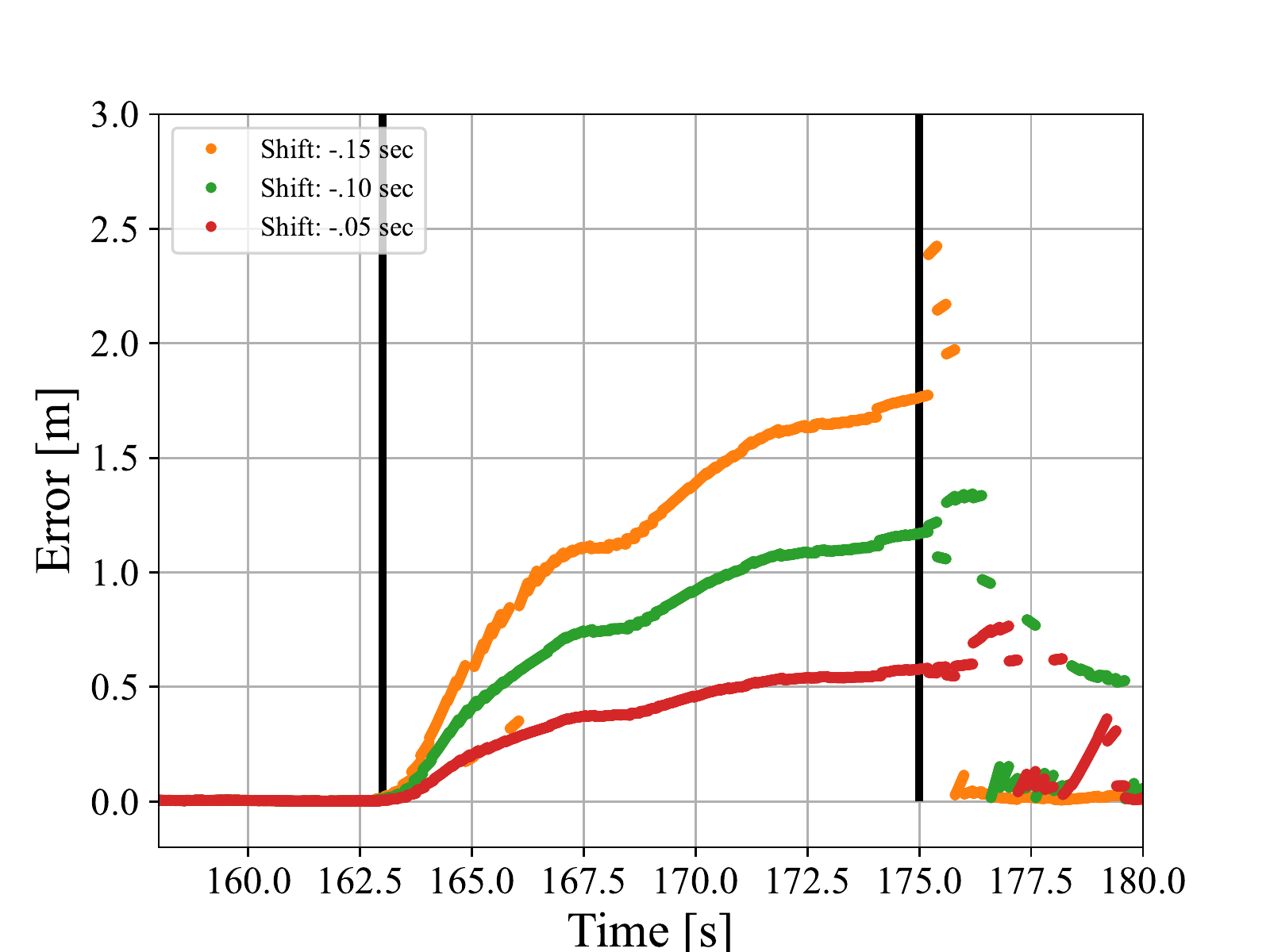}
	\end{minipage}
	\caption{The position errors induced from the different spoofing attacks.  The plot on the left is with the LORD MicroStrain 3DM-GX5-25 (industrial grade) IMU and the plot on the right is with the Bosch BMX055 (consumer grade) IMU. }
	\label{fig:pos_error}
\end{figure}

The corresponding position errors in each attack is shown in Fig.
\ref{fig:pos_error}.  The worst-case attack (time shift of -.05 seconds) only
introduces a .5 meter offset over 10 seconds, indicative of an extremely subtle
attack.  Even the least subtle attack (time shift of -.15 seconds) only
introduces a 2 meter offset after 10 seconds, which is much more challenging
than the attacks simulated in the related work.

\section*{Future Work}
The results from this paper are promising.  It would be beneficial to collect
even more data with the \emph{Sensorium} to strengthen the empirical model.
Future work includes developing techniques for the vehicle to be ``contextually
aware" of what multipath environment it resides in.  This could be achieved by
monitoring certain heuristics such as SNR, tracking loop status, and the signal
quality from low elevation satellites.  It would also be useful to derive the
true distribution of carrier-phase fixed-ambiguity residual cost.  Additionally, 
a direct comparison between position offset and timestamp spoofing 
would be insightful.  

\section*{Conclusion}
A powerful single-antenna carrier-phase-based test to detect GNSS spoofing
attacks on ground vehicles equipped with a low-cost IMU was developed,
implemented, and validated.  Artificial worst-case spoofing attacks were
injected into a dataset collected by a vehicle-mounted sensor suite in Austin,
Texas and detected within two seconds.  This was accomplished by using a
spoofing detection technique that capitalized on the carrier phase fixed-ambiguity residual cost
produced by a well-calibrated CDGNSS solution that is tightly coupled with a
low-cost IMU. The finer movements of the vehicle, such as slight steering
movements and road vibrations, are the necessary unpredictable dithering a
spoofer is not able to replicate.  The differences between
IMU-predicted and measured carrier phase values offer the basis for an exquisitely 
sensitive GNSS spoofing detection statistic.  This paper developed the null-hypothesis 
empirical distributions for the test statistic in both shallow and deep urban areas, and
uses these distributions to demonstrate that high-sensitivity spoofing
detection is possible despite integer folding and urban multipath.
Additionally, the effectiveness of consumer- and industrial-grade IMUs for
spoofing detection was compared.  The type of tightly-coupled IMU-GNSS
estimator whose by-products the proposed detection technique exploits is not
currently available on commercial passenger vehicles, but can be expected to be
adopted in future automated vehicles, since it provides all-weather dm-level
absolute positioning.

\section*{Acknowledgments}
This work was supported in part by the U.S. Department of Transportation
(USDOT) under Grant 69A3552047138 for the CARMEN University Transportation
Center (UTC), and by the Army Research Office under Cooperative Agreement
W911NF-19-2-0333. The views and conclusions contained in this document are
those of the authors and should not be interpreted as representing the official
policies, either expressed or implied, of the Army Research Office or the
U.S. Government. The U.S. Government is authorized to reproduce and distribute
reprints for Government purposes notwithstanding any copyright notation herein.

\bibliographystyle{aiaa} 
\bibliography{ms}

\begin{thebibliography}{10}
\newcommand{\enquote}[1]{``#1''}

\bibitem{volpe2001gps}
{John A. Volpe National Transportation Systems Center}, \enquote{Vulnerability
  assessment of the transportation infrastructure relying on the {Global
  Positioning System},} 2001.

\bibitem{psiakiNewBlueBookspoofing}
Psiaki, M.~L. and Humphreys, T.~E., {\em Position, Navigation, and Timing
  Technologies in the 21st Century: Integrated Satellite Navigation, Sensor
  Systems, and Civil Applications\/}, Vol.~1, chap. {C}ivilian {GNSS}
  {S}poofing, {D}etection, and {R}ecovery, Wiley-IEEE, 2020, pp. 655--680.

\bibitem{mit2020TeslaSpoof}
Mit, R., Zangvil, Y., and Katalan, D., \enquote{Analyzing Tesla‘s Level 2
  Autonomous Driving System Under Different GNSS Spoofing Scenarios and
  Implementing Connected Services for Authentication and Reliability of GNSS
  Data,} {\em Proceedings of the 33rd International Technical Meeting of the
  Satellite Division of The Institute of Navigation (ION GNSS+ 2020)\/}, 2020,
  pp. 621--646.

\bibitem{psiaki2016gnssSpoofing}
Psiaki, M.~L. and Humphreys, T.~E., \enquote{{GNSS} Spoofing and Detection,}
  {\em Proceedings of the IEEE\/}, Vol.~104, No.~6, 2016, pp.~1258--1270.

\bibitem{wesson2018pincer}
Wesson, K.~D., Gross, J.~N., Humphreys, T.~E., and Evans, B.~L.,
  \enquote{{GNSS} Signal Authentication Via Power and Distortion Monitoring,}
  {\em IEEE Transactions on Aerospace and Electronic Systems\/}, Vol.~54,
  No.~2, April 2018, pp.~739--754.

\bibitem{gross2018maximum}
Gross, J.~N., Kilic, C., and Humphreys, T.~E., \enquote{Maximum-likelihood
  power-distortion monitoring for {GNSS}-signal authentication,} {\em IEEE
  Transactions on Aerospace and Electronic Systems\/}, Vol.~55, No.~1, 2018,
  pp.~469--475.

\bibitem{humphreysGNSShandbook}
Humphreys, T.~E., \enquote{Interference,} {\em Springer Handbook of Global
  Navigation Satellite Systems\/}, Springer International Publishing, 2017, pp.
  469--503.

\bibitem{montgomery2009pcgs}
Montgomergy, P.~Y., Humphreys, T.~E., and Ledvina, B.~M.,
  \enquote{Receiver-Autonomous Spoofing Detection: {E}xperimental Results of a
  Multi-antenna Receiver Defense Against a Portable Civil {GPS} Spoofer,} {\em
  Proceedings of the ION International Technical Meeting\/}, Anaheim, CA, Jan.
  2009.

\bibitem{psiaki2014wrod}
Psiaki, M.~L., O'Hanlon, B.~W., Powell, S.~P., Bhatti, J.~A., Wesson, K.~D.,
  Humphreys, T.~E., and Schofield, A., \enquote{{GNSS} Spoofing Detection using
  Two-Antenna Differential Carrier Phase,} {\em Proceedings of the {ION}
  {GNSS}+ Meeting\/}, Institute of Navigation, Tampa, FL, 2014.

\bibitem{reid2019localization}
Reid, T.~G., Houts, S.~E., Cammarata, R., Mills, G., Agarwal, S., Vora, A., and
  Pandey, G., \enquote{Localization Requirements for Autonomous Vehicles,} {\em
  arXiv preprint arXiv:1906.01061\/}, 2019.

\bibitem{ye2019tightly}
Ye, H., Chen, Y., and Liu, M., \enquote{Tightly coupled 3d lidar inertial
  odometry and mapping,} {\em 2019 International Conference on Robotics and
  Automation (ICRA)\/}, IEEE, 2019, pp. 3144--3150.

\bibitem{chiang2020navigation}
Chiang, K.-W., Tsai, G.-J., Li, Y.-H., Li, Y., and El-Sheimy, N.,
  \enquote{Navigation Engine Design for Automated Driving Using {INS/GNSS/3D
  LiDAR-SLAM} and Integrity Assessment,} {\em Remote Sensing\/}, Vol.~12,
  No.~10, 2020, pp.~1564.

\bibitem{narula2021radarpositioningjournal}
Narula, L., Iannucci, P.~A., and Humphreys, T.~E., \enquote{Towards all-weather
  sub-50-cm radar-inertial positioning,} {\em Field Robotics\/}, 2021, To be
  published{.}

\bibitem{yoder2022tightCoupling}
Yoder, J.~E. and Humphreys, T.~E., \enquote{Low-Cost Inertial Aiding for
  Deep-Urban Tightly-Coupled Multi-Antenna Precise GNSS,} {\em Navigation,
  Journal of the Institute of Navigation\/}, 2022, Submitted for review{.}

\bibitem{teunissenGNSShandbookCdgnss}
Teunissen, P. J.~G., {\em Springer Handbook of Global Navigation Satellite
  Systems\/}, chap. Carrier Phase Integer Ambiguity Resolution, Springer, 2017,
  pp. 661--685.

\bibitem{humphreys2019deepUrbanIts}
Humphreys, T.~E., Murrian, M.~J., and Narula, L., \enquote{Deep-Urban Unaided
  Precise Global Navigation Satellite System Vehicle Positioning,} {\em {IEEE}
  Intelligent Transportation Systems Magazine\/}, Vol.~12, No.~3, 2020,
  pp.~109--122.

\bibitem{khanafseh2014raim}
Khanafseh, S., Roshan, N., Langel, S., Cheng-Chan, F., Joerger, M., and Pervan,
  B., \enquote{{GPS} Spoofing Detection Using {RAIM} with {INS} Coupling,} {\em
  Proceedings of the IEEE/ION PLANS Meeting\/}, May 2014.

\bibitem{tanil2017detecting}
Tan{\i}l, {\c{C}}., Khanafseh, S., and Pervan, B., \enquote{Detecting Global
  Navigation Satellite System Spoofing Using Inertial Sensing of Aircraft
  Disturbance,} {\em Journal of Guidance, Control, and Dynamics\/}, 2017.

\bibitem{tanil2018insMonitor}
Tanil, C., Khanafseh, S., Joerger, M., and Pervan, B., \enquote{An INS Monitor
  to Detect {GNSS} Spoofers Capable of Tracking Vehicle Position,} {\em IEEE
  Transactions on Aerospace and Electronic Systems\/}, Vol.~54, No.~1, Feb.
  2018, pp.~131--143.

\bibitem{tanil2018experimental}
Tanil, C., Jimenez, P.~M., Raveloharison, M., Kujur, B., Khanafseh, S., and
  Pervan, B., \enquote{Experimental validation of INS monitor against GNSS
  spoofing,} {\em Proceedings of the 31st International Technical Meeting of
  the Satellite Division of The Institute of Navigation (ION GNSS+ 2018)\/},
  2018, pp. 2923--2937.

\bibitem{tanil2018sequential}
Tanil, C., Khanafseh, S., Joerger, M., and Pervan, B., \enquote{Sequential
  integrity monitoring for Kalman filter innovations-based detectors,} {\em
  Proceedings of the 31st International Technical Meeting of the Satellite
  Division of The Institute of Navigation (ION GNSS+ 2018)\/}, 2018, pp.
  2440--2455.

\bibitem{kujur2020solution}
Kujur, B., Khanafseh, S., and Pervan, B., \enquote{A {S}olution {S}eparation
  {M}onitor using {INS} for {D}etecting {GNSS} {S}poofing,} {\em Proceedings of
  the 33rd International Technical Meeting of the Satellite Division of The
  Institute of Navigation (ION GNSS+ 2020)\/}, 2020, pp. 3210--3226.

\bibitem{liu2019analysis}
Liu, Y., Li, S., Fu, Q., Liu, Z., and Zhou, Q., \enquote{Analysis of Kalman
  filter innovation-based GNSS spoofing detection method for INS/GNSS
  integrated navigation system,} {\em IEEE Sensors Journal\/}, Vol.~19, No.~13,
  2019, pp.~5167--5178.

\bibitem{curran2017use}
Curran, J.~T. and Broumendan, A., \enquote{On the use of low-cost IMUs for GNSS
  spoofing detection in vehicular applications,} {\em Proc. ITSNT\/}, 2017, pp.
  1--8.

\bibitem{psiaki2005relative}
Psiaki, M.~L. and Mohiuddin, S., \enquote{Relative navigation of high-altitude
  spacecraft using dual-frequency civilian {CDGPS},} {\em Proceedings of the
  18th International Technical Meeting of the Satellite Division of The
  Institute of Navigation (ION GNSS 2005)\/}, 2005, pp. 1191--1207.

\bibitem{psiakiAmbiguity2007}
Psiaki, M. and Mohiuddin, S., \enquote{Global Positioning System Integer
  Ambiguity Resolution Using Factorized Least-Squares Techniques,} {\em Journal
  of Guidance, Control, and Dynamics\/}, Vol.~30, No.~2, March-April 2007,
  pp.~346--356.

\bibitem{s_mohiuddin07_wia}
Mohiuddin, S. and Psiaki, M.~L., \enquote{High-Altitude Satellite Relative
  Navigation Using Carrier-Phase Differential Global Positioning System
  Techniques,} {\em Journal of Guidance, Control, and Dynamics\/}, Vol.~30,
  No.~5, Sept.-Oct. 2007, pp.~1628--1639.

\bibitem{narula2020texcup}
Narula, L., LaChapelle, D.~M., Murrian, M.~J., Wooten, J.~M., Humphreys, T.~E.,
  Lacambre, J.-B., de~Toldi, E., and Morvant, G., \enquote{{TEX-CUP}: {T}he
  {U}niversity of {T}exas {C}hallenge for {U}rban {P}ositioning,} {\em
  Proceedings of the IEEE/ION PLANSx Meeting\/}, 2020.

\bibitem{t_humphreys06_scp}
Humphreys, T.~E., Ledvina, B.~M., Psiaki, M.~L., and Kintner, Jr., P.~M.,
  \enquote{{GNSS} Receiver Implementation on a {DSP}: Status, Challenges, and
  Prospects,} {\em Proceedings of the ION GNSS Meeting\/}, Institute of
  Navigation, Fort Worth, TX, 2006, pp. 2370--2382.

\bibitem{lightsey2013demonstration}
Lightsey, E.~G., Humphreys, T.~E., Bhatti, J.~A., Joplin, A.~J., O'Hanlon,
  B.~W., and Powell, S.~P., \enquote{Demonstration of a Space Capable Miniature
  Dual Frequency {GNSS} Receiver,} {\em Navigation\/}, Vol.~61, No.~1, Mar.
  2014, pp.~53--64.

\bibitem{t_humphreys09_sdgr}
Humphreys, T.~E., Bhatti, J., Pany, T., Ledvina, B., and O'Hanlon, B.,
  \enquote{Exploiting multicore technology in software-defined {GNSS}
  receivers,} {\em Proceedings of the ION GNSS Meeting\/}, Institute of
  Navigation, Savannah, GA, 2009, pp. 326--338.

\bibitem{yoder2020visonFusion}
Yoder, J.~E., Iannucci, P.~A., Narula, L., and Humphreys, T.~E.,
  \enquote{Multi-Antenna Vision-and-Inertial-Aided {CDGNSS} for Micro Aerial
  Vehicle Pose Estimation,} {\em Proceedings of the {ION} {GNSS}+ Meeting\/},
  Online, 2020.

\bibitem{clements2021bitpackingIonGnss}
Clements, Z., Iannucci, P.~A., Humphreys, T.~E., and Pany, T.,
  \enquote{Optimized Bit-Packing for Bit-Wise Software-Defined {GNSS} Radio,}
  {\em Proceedings of the {ION} {GNSS}+ Meeting\/}, St. Louis, MO, 2021.

\bibitem{kerns2014unmanned}
Kerns, A.~J., Shepard, D.~P., Bhatti, J.~A., and Humphreys, T.~E.,
  \enquote{Unmanned Aircraft Capture and Control Via {GPS} Spoofing,} {\em
  Journal of Field Robotics\/}, Vol.~31, No.~4, 2014, pp.~617--636.

\bibitem{shepard2012droneHack}
Shepard, D.~P., Bhatti, J.~A., and Humphreys, T.~E., \enquote{Drone Hack:
  Spoofing Attack Demonstration on a Civilian Unmanned Aerial Vehicle,} {\em
  {GPS} World\/}, Aug. 2012.

\bibitem{bhatti2017hostile}
Bhatti, J. and Humphreys, T.~E., \enquote{Hostile control of ships via false
  {GPS} signals: Demonstration and detection,} {\em Navigation\/}, Vol.~64,
  No.~1, 2017, pp.~51--66.

\bibitem{t_humphreys2014stb}
Humphreys, T.~E., Shepard, D.~P., Bhatti, J.~A., and Wesson, K.~D., \enquote{A
  Testbed for Developing and Evaluating {GNSS} Signal Authentication
  Techniques,} {\em Proceedings of the International Symposium on Certification
  of {GNSS} Systems and Services (CERGAL)\/}, Dresden, Germany, July 2014.

\bibitem{humphreys2012_TEST_Battery}
Humphreys, T.~E., Bhatti, J.~A., Shepard, D.~P., and Wesson, K.~D.,
  \enquote{The {Texas Spoofing Test Battery}: Toward a Standard for Evaluating
  {GNSS} Signal Authentication Techniques,} {\em Proceedings of the ION GNSS
  Meeting\/}, 2012.

\bibitem{rnlTexbatSite}
Laboratory, T.~R., \enquote{{Texas Spoofing Test Battery (TEXBAT)},} July 2017,
  \url{http://radionavlab.ae.utexas.edu/texbat}.

\end{thebibliography}

\end{document}